\documentclass[preprint,prd,showpacs,showkeys]{revtex4-1}

\usepackage{amsmath}
\usepackage{amssymb}
\usepackage{amsfonts}
\usepackage{graphicx}
\usepackage{axodraw4j}
\usepackage{appendix}

\begin{document}
\title{Gauge Invariant Treatment of $\gamma_{5}$ in the Scheme of 't Hooft and Veltman}
\author{Er-Cheng Tsai}
\email{ectsai@ntu.edu.tw}
\affiliation{Physics Department, National Taiwan University, Taipei, Taiwan}

\begin{abstract}
We propose moving all the $\gamma_{5}$ matrices to the rightmost position
before continuing the dimension, and show that this simple prescription will
enable the dimension regularization scheme proposed by 't Hooft and Veltman to
be consistent with gauge invariance.
\end{abstract}

\maketitle

\section{Introduction}

The dimensional regularization scheme of 't Hooft and Veltman \cite{HV} has
proven to be successful\ not only for its theoretical implication in
renormalizing gauge field theories but also for its practical application in
simplifying the calculations of Feynman amplitudes. Only one difficulty of the
dimensional regularization scheme remains. As pointed out by 't Hooft and
Veltman in Sec. 6 of \cite{HV}, the breakdown of Ward identities \cite{WTI} in
the presence of
\begin{equation}
\gamma_{5}=i\gamma^{0}\gamma^{1}\gamma^{2}\gamma^{3} \label{g5def}%
\end{equation}
is caused by the inability of maintaining the validity of the basics identity%
\begin{equation}
\not k  \gamma_{5}=\left(  \not p  +\not k  \right)  \gamma_{5}+\gamma
_{5}\not p  \label{w1}%
\end{equation}
beyond dimension $n=4$. In this paper, we shall present a simple method to
deal with $\gamma_{5}$ which reserves the validity of the basic Ward identity
$\left(  \ref{w1}\right)  $ so that the dimensional regularization scheme can
still be useful in yielding gauge invariant regularized amplitudes for gauge
theories involving $\gamma_{5}$.

In a four-dimensional space, any matrix product
\[
\hat{M}=\gamma_{\omega_{1}}\gamma_{\omega_{2}}...\gamma_{\omega_{n}}\text{
with }\omega_{i}\in\left\{  0,1,2,3,5\right\}
\]
may be reduced, by anti-commuting $\gamma_{5}$ to the rightmost position, to
either the form of $\pm\gamma_{\mu_{1}}\gamma_{\mu_{2}}...\gamma_{\mu_{m}}$
with $\mu_{i}\in\left\{  0,1,2,3\right\}  $ if $\hat{M}$ contains even
$\gamma_{5}$ factors, or the form $\pm\gamma_{\nu_{1}}\gamma_{\nu_{2}%
}...\gamma_{\nu_{p}}\gamma_{5}$ with $\nu_{i}\in\left\{  0,1,2,3\right\}  $ if
the $\gamma_{5}$ count is odd. The matrix product $\gamma_{\mu_{1}}\gamma
_{\mu_{2}}...\gamma_{\mu_{m}}$ is unambiguously defined under dimensional
regularization when the components $\mu_{i}$ run out of the range $\left\{
0,1,2,3\right\}  $ of the first four dimensions. The product $\gamma_{\nu_{1}%
}\gamma_{\nu_{2}}...\gamma_{\nu_{p}}\gamma_{5}$ with one $\gamma_{5}$ on the
right is continued by defining it to be the product of the analytically
continued $\gamma_{\nu_{1}}\gamma_{\nu_{2}}...\gamma_{\nu_{p}}$ and the
$\gamma_{5}$ in $\left(  \ref{g5def}\right)  $. Before analytic continuation
is made, a $\gamma_{5}$-odd matrix product may always be reduced to a matrix
product with only one $\gamma_{5}$ factor. To analytically continue such a
matrix product, we adopt the default continuation by anti-commuting the
$\gamma_{5}$ matrix to the rightmost position before making the continuation.

Let us introduce the notation $\underline{p^{\mu}}$ for the component of
$p^{\mu}$ vector in the first 4 dimensions and the notation $p_{\Delta}^{\mu}$
for the component in the remaining $n-4$ dimensions. $i.e.$, $p^{\mu
}=\underline{p}^{\mu}+p_{\Delta}^{\mu}$ with $p_{\Delta}^{\mu}=0$\ if $\mu
\in\left\{  0,1,2,3\right\}  $ and $\underline{p}^{\mu}=0$\ if $\mu
\notin\left\{  0,1,2,3\right\}  $. The Dirac matrix $\gamma^{\mu}$ is also
decomposed as $\gamma^{\mu}=\underline{\gamma}^{\mu}+\gamma_{\Delta}^{\mu}$
with $\gamma_{\Delta}^{\mu}=0$\ when $\mu\in\left\{  0,1,2,3\right\}  $ and
$\underline{\gamma}^{\mu}=0$ when $\mu\notin\left\{  0,1,2,3\right\}  $. We
then have%
\begin{equation}
\gamma_{5}\gamma^{\mu}+\gamma^{\mu}\gamma_{5}=2\gamma_{\Delta}^{\mu}\gamma
_{5}, \label{g5ant2}%
\end{equation}
which means that $\gamma_{5}$\ does not anti-commute with $\gamma^{\mu}$ when
$\mu\notin\left\{  0,1,2,3\right\}  $.

For the QED theory, the identity%
\begin{equation}
\frac{1}{\not \ell -m}\not k  \frac{1}{\not \ell -\not k  -m}=\frac{1}%
{\ell-\not k  -m}-\frac{1}{\not \ell -m} \label{qedwti0}%
\end{equation}
is the foundation that a Ward identity is built upon. For a gauge theory
involving $\gamma_{5}$, there is a basic identity similar to $\left(
\ref{qedwti0}\right)  $ for verifying Ward identities:%
\begin{equation}
\frac{1}{\not \ell +\not k  -m}\left(  \not k  -2m\right)  \gamma_{5}\frac
{1}{\not \ell -m}=\gamma_{5}\frac{1}{\not \ell -m}+\frac{1}{\not \ell +\not k
-m}\gamma_{5} \label{canc}%
\end{equation}
The above identity valid at $n=4$ is derived by decomposing the vertex factor
$\left(  \not k  -2m\right)  \gamma_{5}$ into $\left(  \not \ell +\not k
-m\right)  \gamma_{5}$ and $\gamma_{5}\left(  \not \ell -m\right)  $ that
annihilate respectively the propagators of the outgoing fermion with momentum
$\ell+k$ and the incoming fermion with momentum $\ell$. Positioning
$\gamma_{5}$ at the rightmost site, the above identity becomes%
\begin{equation}
\frac{1}{\not \ell +\not k  -m}\left(  \not k  -2m\right)  \frac
{1}{-\not \ell -m}\gamma_{5}=\left(  \frac{1}{-\not \ell -m}+\frac
{1}{\not \ell +\not k  -m}\right)  \gamma_{5} \label{bt5}%
\end{equation}
If we disregard the rightmost $\gamma_{5}$ on both sides of the above
identity, we obtain another identity that is valid at $n=4$. This new
identity, which is void of $\gamma_{5}$, may be analytically continued to hold
when $n\neq4$. We then multiply $\gamma_{5}$ on the right to every
analytically continued term of this $\gamma_{5}$-free identity to yield the
analytic continuation of the identity $\left(  \ref{canc}\right)  $.

As a side remark, we note that when we go to the dimension of $n\neq4$,
$\left(  \ref{canc}\right)  $ in the form presented above is not valid. This
is because $\gamma_{5}$ defined in $\left(  \ref{g5def}\right)  $ does not
always anti-commute with $\gamma^{\mu}$ if $n\neq4$. Adopting the rightmost
$\gamma_{5}$ ordering avoids this difficulty, as the validity of the identity
in the form of rightmost $\gamma_{5}$ ordering no longer depends on
$\gamma_{5}$ anti-commuting with the $\gamma$ matrices.

For an amplitude corresponding to a diagram involving no fermion loops, we
shall move all $\gamma_{5}$ matrices to the rightmost position before we
continue analytically the dimension $n$. Subsequent application of dimensional
regularization gives us regulated amplitudes satisfying the Ward identities.

If a diagram has one or more fermion loops, the amplitude corresponding to
this diagram can be regulated in more than one ways. This is because there are
different ways to assign the starting position on a fermion loop. Once we have
chosen a starting point, we define the matrix product inside the trace by
rightmost $\gamma_{5}$ ordering before making the analytic continuation. In
general, continuations from different starting points give different values
for the trace when $n\neq4$. An identity relating the traces of matrix
products without $\gamma_{5}$ at $n=4$ can always be analytically continued to
hold when $n\neq4$. Therefore, the portion of an amplitude in which the count
of $\gamma_{5}$ on every loop is even has no $\gamma_{5}$ difficulty
\cite{GD}. But to calculate amplitudes with an odd count of $\gamma_{5}$, we
need an additional prescription. This is because, as we have mentioned, the
rightmost position on a fermion loop is not defined a-priori.

Although we have multiple continuations for the trace of a matrix product,
they differ with one another by terms containing at least a factor of
$\gamma_{\Delta}^{\mu}$. In the tree order and in the limit $n\rightarrow4$,
they are all restored to the same result because $\gamma_{\Delta}^{\mu}$ will
disappear when $n\rightarrow4$. For higher loop orders, $\gamma_{\Delta}^{\mu
}$ contribution may not be ignored. This is because the factor $\gamma
_{\Delta}^{\mu}\gamma_{\Delta}^{\nu}g_{\mu\nu}=\left(  n-4\right)  $
multiplied by a divergent integral, which generates a simple pole factor
$\frac{1}{\left(  n-4\right)  }$ or a higher-order pole term, is finite or
even infinite in the limit $n\rightarrow4$. Thus divergent diagrams with
fermion loops are the only type of diagrams that may be ambiguous with respect
to the $\gamma_{5}$ positioning. Not incidentally, they are also the diagrams
that may be plagued by anomaly problem \cite{ABJ}.

In the dimensional regularization scheme of Breitenlohner and Maison
\cite{BM}, the $\gamma_{5}$ matrix is also defined as $\left(  \ref{g5def}%
\right)  $. By continuing the Lagrangian to dimension $n\neq4$ and carefully
handling the evanescent terms that are proportional to $\left(  n-4\right)  $,
Breitenlohner and Maison were able to show that dimensional regularization and
minimal-subtraction renormalization can be implemented consistently for
theories involving $\gamma_{5}$. But there is a major deficiency of this BM
scheme: it is not a gauge invariant scheme. Consequently, amplitudes obtained
therewith do not satisfy Ward identities and finite counter terms are required
to restore the validities of these identities \cite{GB2,AT,FER,TLTM,FG,SANR}.
This in fact renders the application of dimensional regularization for chiral
gauge theories rather complicated in practical calculation.

There is a naive dimensional regularization (NDR) scheme which assumes that
$\gamma_{5}$ satisfies $\gamma_{5}\gamma^{\mu}+\gamma^{\mu}\gamma_{5}=0$ for
all $\mu$ even when $n\neq4$. Since no such $\gamma_{5}$\ exists, this scheme
is not without fault. In particular, it is not capable of producing the
triangular anomaly term. While regulated amplitudes satisfying Ward identities
have often been obtained in the past with the use of the NDR scheme
\cite{CFH,NDR-EXS,BA etc}, it is because all the $\gamma_{5}$ matrices have
been tacitly moved outside of divergent sub-diagrams in these calculations.
This is to say that the rightmost $\gamma_{5}$ scheme has been employed in actuality.

In contrast to the NDR scheme, the triangular diagrams that are responsible
for the anomaly can be handled by the rightmost $\gamma_{5}$ scheme. We shall
find that, while there are many choices for the rightmost position on a loop,
some choices are in violation of gauge invariance.\ When no choice obeying all
symmetry requirements is available, the Ward identity involving such
amplitudes may be broken to give rise to an anomaly.

K\"{o}rner, Kreimer and Schilcher \cite{KKS} have shown that if we relinquish
the cyclicity of the trace for a matrix product, an anticommuting $\gamma_{5}$
as the one adopted in the NDR scheme can be defined under dimensional
regularization. In this KKS formalism, the non-cyclic trace becomes the
spoiler of Ward identities because the trace of a matrix product with an odd
number of $\gamma_{5}$ depends on where the `reading point' is designated. For
a fermion loop, the reading point in the KKS scheme plays the similar role as
the rightmost $\gamma_{5}$ posistion in our scheme. In fact, they both yield
the same dimensionally regularized amplitudes for the fermion loop if the
reading point and the rightmost $\gamma_{5}$ are identically positioned. But
the prescription for choosing reading point given in \cite{KKS} to resolve the
non-cyclicity difficulty is flawed because the effect of the pole terms
arising from loop integrals of divergent sub-diagrams or overall diagram on
the $O\left(  n-4\right)  $\ difference due to different reading points has
been mistakenly ignored. As a consequence, the KKS method presented in
\cite{KKS} does not lead to regularized amplitudes that obey multi-loop Ward
identities. This problem can be solved, as we shall verify in this paper, by
using only the reading points or $\gamma_{5}$ positions that are located
outside all self-energy-insertion or vertex-correction sub-diagrams and are
not at any vertices connecting to external field lines.

By taking advantage of the rightmost $\gamma_{5}$ scheme, which allows us to
employ identities such as $\left(  \ref{w1}\right)  $ and $\left(
\ref{canc}\right)  $\ given above, we will be able to choose rightmost
positions for the $\gamma_{5}$ to be in consistency with gauge invariance so
that Ward identities are preserved. In particular, we shall demonstrate how to
apply the rightmost $\gamma_{5}$ scheme to account for the 1-loop anomaly and
the preservation of the 2-loop triangular Ward identity in the chiral
Abelian-Higgs theory defined below in the following section. This Abelian
theory has no infrared problem and has a non-free ghost field due to the
particular gauge fixing term that we choose to use. The $\gamma_{5}$ treatment
that we shall present for the Abelian theory is also applicable to non-Abelian
theories which must be accompanied by ghost interactions.

\section{Abelian-Higgs Gauge Theory with Chiral Fermion}

The Lagrangian for the Abelian-Higgs gauge theory \cite{HIG} with chiral
fermion is%
\begin{align}
L &  =-{\frac{1}{4}}F_{\mu\nu}F^{\mu\nu}+\left(  D^{\mu}\phi\right)
^{\dagger}\left(  D_{\mu}\phi\right)  -\frac{1}{2}\lambda g^{2}\left(
\phi^{\dagger}\phi-\frac{1}{2}v^{2}\right)  ^{2}\label{e2-1}\\
&  +\bar{\psi}_{L}\left(  i\not D\right)  \psi_{L}+\bar{\psi}_{R}\left(
i\not \partial\right)  \psi_{R}-\sqrt{2}f\left(  \bar{\psi}_{L}\phi\psi
_{R}+\bar{\psi}_{R}\phi^{\dagger}\psi_{L}\right)  ,\nonumber
\end{align}
where%
\[
F_{\mu\nu}\equiv\partial_{\mu}A_{\nu}-\partial_{\nu}A_{\mu},
\]%
\[
D_{\mu}\phi\equiv\left(  \partial_{\mu}+igA_{\mu}\right)  \phi,
\]%
\[
\psi_{L}=L\psi,\,\psi_{R}=R\psi
\]
with the chiral projection operators $L$ and $R$ defined as%
\[
L=\frac{1}{2}\left(  1-\gamma_{5}\right)  ,R=\frac{1}{2}\left(  1+\gamma
_{5}\right)  .
\]
We define two Hermitian fields $H$ and $\phi_{2}$ for the real and imaginary
parts of the complex scalar field by%
\begin{equation}
\phi=\dfrac{H+i\phi_{2}+v}{\sqrt{2}}.\label{defphi}%
\end{equation}
We also introduce two mass parameters $M$ and $m$ defined by%
\begin{equation}
M=gv,m=fv\label{mdef}%
\end{equation}
Both $M$ and $m$ will be regarded as zero order quantities in perturbation. To
quantize this theory, we add to the Lagrangian $L$ gauge fixing terms as well
as the associated ghost terms. The sum will be called the effective Lagrangian
$L_{eff}$, and is invariant under the following BRST \cite{BRS,CTBRS}
variations:%
\begin{align}
\delta A_{\mu} &  =\partial_{\mu}c,\label{brst}\\
\delta\phi_{2} &  =-Mc-gcH,\nonumber\\
\delta H &  =gc\phi_{2},\nonumber\\
\delta\psi_{L} &  =-igc\psi_{L},\delta\psi_{R}=0,\nonumber\\
\delta\bar{c} &  =-\dfrac{i}{\alpha}\left(  \partial^{\mu}A_{\mu}%
-\alpha\Lambda\phi_{2}\right)  ,\delta c=0.\nonumber
\end{align}
where $c$ is the ghost field and $\bar{c}$ is the anti-ghost field. The gauge
fixing term is
\begin{equation}
L_{gf}=-\frac{1}{2\alpha}\left(  \partial_{\mu}A^{\mu}-\alpha\Lambda\phi
_{2}\right)  ^{2}\label{egf1}%
\end{equation}
and the ghost term is
\begin{equation}
L_{ghost}=i\bar{c}\delta\left(  \partial_{\mu}A^{\mu}-\alpha\Lambda\phi
_{2}\right)  =i\bar{c}\left(  \partial_{\mu}\partial^{\mu}+\alpha\Lambda
M\right)  c+ig\alpha\Lambda\bar{c}cH.\label{egh1}%
\end{equation}
The BRST invariant effective Lagrangian is%
\begin{equation}
L_{eff}=L+L_{gf}+L_{ghost}\label{e2-3}%
\end{equation}

\subsection{Graphical Identities}

The prescription of the rightmost $\gamma_{5}$ ordering under dimensional
regularization offers a scheme to construct amplitudes when $n\neq4$. We now
introduce some graphical notations for verifying diagrammatically if the
regularized amplitudes so obtained satisfy Ward identities.

According to the Feynman rules, one assigns the factor $-igR\gamma^{\mu}L$ to
the vertex $\bar{\psi}-A^{\mu}-\psi$ and the factor $-f\left(  L-R\right)  $
to the vertex $\bar{\psi}-\phi_{2}-\psi$, as these factors correspond to the
terms $-g\bar{\psi}_{L}\not A  \psi_{L}$ and $-if\left(  \bar{\psi}_{L}%
\phi_{2}\psi_{R}-\bar{\psi}_{R}\phi_{2}\psi_{L}\right)  $ in the interaction
Lagrangian of $\left(  \ref{e2-1}\right)  $. Let us define the following two
graphical notations for these two vertices:%
\begin{equation}%
\raisebox{-12pt}{
\begin{picture}(18,21) (17,-11)
\SetWidth{0.5}
\SetColor{Black}
\Text(32,-5)[]{\normalsize{\Black{$\mu$}}}
\SetWidth{0.5}
\Line(28.001,8.999)(35.999,1.001)\Line(35.999,8.999)(28.001,1.001)
\end{picture}
}%
=-igR\gamma^{\mu}L\text{, }%
\raisebox{-1pt}{
\begin{picture}(6,8) (28,-12)
\SetWidth{0.5}
\SetColor{Black}
\Vertex(32,-8){3}
\end{picture}
}
=-f\left(  L-R\right)  . \label{defvx}%
\end{equation}
We also introduce the notation%
\begin{equation}%
\raisebox{-14pt}{
\begin{picture}(20,24) (17,-26)
\SetWidth{0.5}
\SetColor{Black}
\COval(32,-8)(5.657,5.657)(45.0){Black}{White}\Line
(29.172,-10.828)(34.828,-5.172)\Line(29.172,-5.172)(34.828,-10.828)
\Text(32,-20)[]{\normalsize{\Black{$k$}}}
\end{picture}
}%
=-gR\not k  L+mg\left(  L-R\right)  \label{defeta}%
\end{equation}
which represents the sum of $-ik_{\mu}$ times the $\bar{\psi}-A^{\mu}-\psi$
vertex factor, with $k$ the momentum of the vector particle flowing into the
vertex and $-M$ times the $\bar{\psi}-\phi_{2}-\psi$ vertex factor. Note that
$Mf$ may be equated to $mg$ according to $\left(  \ref{mdef}\right)  $. The
identity
\begin{equation}
R\not k  L-m\left(  L-R\right)  =\left(  \not \ell +\not k  -m\right)
L-R\left(  \not \ell -m\right)  , \label{bbwti}%
\end{equation}
valid in a four-dimensional space, will be our building block for verifying
various Ward identities involving fermion lines. Indeed, if we set $L=R=1$,
the identity above becomes the familiar identity used in verifying Ward
identities in QED.

Sandwiching equation $\left(  \ref{bbwti}\right)  $ between two fermion
propagators, we get%
\begin{equation}
\frac{1}{\left(  \not \ell +\not k  -m\right)  }\left(  R\not k  L-m\left(
L-R\right)  \right)  \frac{1}{\left(  \not \ell -m\right)  }=L\frac{1}{\left(
\not \ell -m\right)  }-\frac{1}{\left(  \not \ell +\not k  -m\right)  }R
\label{bbi5}%
\end{equation}
Note the similarity of this identity with its familiar counterpart $\left(
\ref{qedwti0}\right)  $\ in QED. As noted before, when we go to the dimension
of $n\neq4$, $\left(  \ref{bbi5}\right)  $ in the form presented above is not
valid. This is because $\gamma_{5}$ does not always anti-commute with
$\gamma^{\mu}$ if $n\neq4$. Adopting the rightmost $\gamma_{5}$ ordering
avoids this difficulty. The above equation multiplied by the coupling constant
$g$ may be expressed graphically as%
\begin{equation}%
\begin{picture}(266,28) (25,-22)
\SetWidth{0.5}
\SetColor{Black}
\Line
[arrow,arrowpos=1,arrowlength=2.5,arrowwidth=1,arrowinset=0.2](256,1)(270,1)
\Line
[arrow,arrowpos=1,arrowlength=2.5,arrowwidth=1,arrowinset=0.2](160,1)(146,1)
\Line(48,-4)(80,-4)
\COval(64,-4)(5.657,5.657)(45.0){Black}{White}\Line
(61.172,-6.828)(66.828,-1.172)\Line(61.172,-1.172)(66.828,-6.828)
\Text(112,-5)[]{\normalsize{\Black{$=$}}}
\Line(240,-4)(272,-4)
\COval(256,-4)(5.657,5.657)(45.0){Black}{White}\Line
(253.172,-6.828)(258.828,-1.172)\Line(253.172,-1.172)(258.828,-6.828)
\Text(208,-4)[]{\normalsize{\Black{$+$}}}
\Text(80,-12)[]{\normalsize{\Black{$\ell$}}}
\Text(40,-12)[]{\normalsize{\Black{$\ell+k$}}}
\Text(160,-16)[]{\normalsize{\Black{$k$}}}
\Text(256,-16)[]{\normalsize{\Black{$k$}}}
\Text(176,-12)[]{\normalsize{\Black{$\ell$}}}
\Text(274,-12)[]{\normalsize{\Black{$\ell$}}}
\Line(144,-4)(176,-4)
\COval(160,-4)(5.657,5.657)(45.0){Black}{White}\Line
(157.172,-6.828)(162.828,-1.172)\Line(157.172,-1.172)(162.828,-6.828)
\end{picture}
\label{gbbwti}%
\end{equation}
where the double line emitting from the composite vertex $%
\begin{picture}(12,12) (10,-8)
\SetWidth{0.5}
\SetColor{Black}
\COval(16,-4)(5.657,5.657)(45.0){Black}{White}\Line
(13.172,-6.828)(18.828,-1.172)\Line(13.172,-1.172)(18.828,-6.828)
\end{picture}%
$ indicates that the fermion propagator is annihilated. In addition, the
double line together with the composite vertex is to be replaced by $-igL$ if
the arrow points to the left, and to be replaced by $igR$ if the arrow points
to the right. Thus the following two diagrams cancel each other if the
corresponding external momenta are the same:%
\begin{equation}%
\begin{picture}(170,16) (15,-22)
\SetWidth{0.5}
\SetColor{Black}
\Line(16,-16)(64,-16)
\Text(80,-16)[]{\normalsize{\Black{$+$}}}
\Line
[arrow,arrowpos=1,arrowlength=2.5,arrowwidth=1,arrowinset=0.2](128,-11)(118,-11)
\Line(96,-16)(144,-16)
\COval(128,-16)(5.657,5.657)(45.0){Black}{White}\Line
(125.172,-18.828)(130.828,-13.172)\Line(125.172,-13.172)(130.828,-18.828)
\Text(168,-16)[]{\normalsize{\Black{$= \ \ 0$}}}
\Line(108.001,-19.999)(115.999,-12.001)\Line(108.001,-12.001)(115.999,-19.999)
\Line(44.001,-19.999)(51.999,-12.001)\Line(44.001,-12.001)(51.999,-19.999)
\Line
[arrow,arrowpos=1,arrowlength=2.5,arrowwidth=1,arrowinset=0.2](32,-11)(42,-11)
\COval(32,-16)(5.657,5.657)(45.0){Black}{White}\Line
(29.172,-18.828)(34.828,-13.172)\Line(29.172,-13.172)(34.828,-18.828)
\end{picture}
\label{wti1}%
\end{equation}
In our convention, the direction of any horizontal fermion line is assumed to
be pointing to the left side unless indicated otherwise.

\subsection{Cut Point}

We note that a fermion loop opens up and becomes a fermion line if we make a
cut at some point on the loop. We shall always choose as the cut point either
the beginning point or the endpoint of an internal fermion line on the loop.
An internal fermion line begins from a vertex and ends at another vertex. When
the cut point is chosen to be the endpoint of an internal fermion line, the
vertex factor is then assigned to appear as the beginning factor and stands at
the right end of the matrix product for the entire open fermion line. And when
the cut point is chosen to be the beginning point of an internal fermion line
that emits from a vertex, the matrix factor corresponding to that vertex will
be assigned to be the terminating factor and stands at the left end of the
matrix product for the entire open fermion line. With the cut point on a
fermion loop chosen and with the fermion loop turned into a fermion line, we
may apply the rule of rightmost ordering for $\gamma_{5}$.

The diagrams in this note are often cut open at the end of an internal fermion
line flowing into a vertex, and as a matter of convenience, we will speak of
such a vertex as the cut point. If the vertex is $\bar{\psi}-H-\psi$ with the
vertex factor $-if$ or $\bar{\psi}-\phi_{2}-\psi$ with the vertex factor
$-f\gamma_{5}$, choosing the cut point to be either the endpoint of the
fermion line flowing into the vertex or the beginning point of the fermion
line leaving the vertex gives us identical rightmost $\gamma_{5}$ positioning
and therefore the same dimensionally regularized amplitude. Furthermore, if
the vertex is $\bar{\psi}-A^{\mu}-\psi$ and the polarization $\mu$ falls
within the first 4 dimensions such that its vertex factor anti-commutes with
$\gamma_{5}$, we also get the same regularized amplitude whether or not the
cut point is in the immediate front or rear of the vertex.

\section{One-Loop Triangular Diagrams\label{1td}}

Let $\Gamma_{F}^{\left(  1\right)  }\left(  A^{\mu},A^{\nu},A^{\rho}%
;k_{1},k_{2},k_{3}\right)  $ denote the 1PI $AAA$ amplitude with one fermion
loop and three external fields $A^{\mu},A^{\nu},A^{\rho}$ , with $k_{1}$,
$k_{2}$ $,$ $k_{3}=-k_{1}-k_{2}$ the momenta of $A^{\mu}$, $A^{\nu},$
$A^{\rho}$ , respectively. We may omit the momentum variables $k_{1}$, $k_{2}$
$,$ $k_{3}$ if there is no confusion. The superscript $^{\left(  1\right)  }$
signifies that the amplitude is of one loop, while the subscript $_{F}$
signifies the presence of a fermion loop. The directions of the external
momenta are inward. Similarly, $\Gamma_{F}^{\left(  1\right)  }\left(  A^{\mu
},A^{\nu},\phi_{2};k_{1},k_{2},k_{3}\right)  $ denotes the 1PI $AA\phi_{2}$
amplitude with one fermion loop. The Ward identity that relates $\Gamma
_{F}^{\left(  1\right)  }\left(  A^{\mu},A^{\nu},A^{\rho}\right)  $ to
$\Gamma_{F}^{\left(  1\right)  }\left(  A^{\mu},A^{\nu},\phi_{2}\right)  $ is%
\begin{equation}
-ik_{3}^{\rho}\Gamma_{F}^{\left(  1\right)  }\left(  A^{\mu},A^{\nu},A^{\rho
};k_{1},k_{2},k_{3}\right)  -M\Gamma_{F}^{\left(  1\right)  }\left(  A^{\mu
},A^{\nu},\phi_{2};k_{1},k_{2},k_{3}\right)  =0. \label{triwti}%
\end{equation}
Formally, amplitude for the left side of the above identity $\left(
\ref{triwti}\right)  $ is represented by the sum of the following two Feynman
diagrams:%
\begin{equation}%
\begin{picture}(128,52) (1,-10)
\SetWidth{0.5}
\SetColor{Black}
\Line
[arrow,arrowpos=0.5,arrowlength=2.5,arrowwidth=1,arrowinset=0.2](80,4)(112,4)
\Line
[arrow,arrowpos=0.5,arrowlength=2.5,arrowwidth=1,arrowinset=0.2](96,36)(80,4)
\Line
[arrow,arrowpos=0.5,arrowlength=2.5,arrowwidth=1,arrowinset=0.2](112,4)(96,36)
\Text(112,-4)[]{\normalsize{\Black{$\mu$}}}
\Text(80,-4)[]{\normalsize{\Black{$\nu$}}}
\COval(96,36)(5.657,5.657)(45.0){Black}{White}\Line
(93.172,33.172)(98.828,38.828)\Line(93.172,38.828)(98.828,33.172)
\Line(108.001,0.001)(115.999,7.999)\Line(108.001,7.999)(115.999,0.001)
\Line(76.001,0.001)(83.999,7.999)\Line(76.001,7.999)(83.999,0.001)
\Line
[arrow,arrowpos=0.5,arrowlength=2.5,arrowwidth=1,arrowinset=0.2](16,4)(48,4)
\Line
[arrow,arrowpos=0.5,arrowlength=2.5,arrowwidth=1,arrowinset=0.2](32,36)(16,4)
\Line
[arrow,arrowpos=0.5,arrowlength=2.5,arrowwidth=1,arrowinset=0.2](48,4)(32,36)
\Text(48,-4)[]{\normalsize{\Black{$\nu$}}}
\Text(16,-4)[]{\normalsize{\Black{$\mu$}}}
\COval(32,36)(5.657,5.657)(45.0){Black}{White}\Line
(29.172,33.172)(34.828,38.828)\Line(29.172,38.828)(34.828,33.172)
\Line(44.001,0.001)(51.999,7.999)\Line(44.001,7.999)(51.999,0.001)
\Line(12.001,0.001)(19.999,7.999)\Line(12.001,7.999)(19.999,0.001)
\end{picture}
\label{a-01}%
\end{equation}

If we replace the circled cross $%
\begin{picture}(12,12) (10,-8)
\SetWidth{0.5}
\SetColor{Black}
\COval(16,-4)(5.657,5.657)(45.0){Black}{White}\Line
(13.172,-6.828)(18.828,-1.172)\Line(13.172,-1.172)(18.828,-6.828)
\end{picture}%
$ in the two diagrams above by the uncircled cross $%
\begin{picture}(12,12) (27,-1)
\SetWidth{0.5}
\SetColor{Black}
\Text(32,-2)[]{\normalsize{\Black{$\rho$}}}
\SetWidth{0.5}
\Line(28.001,8.999)(35.999,1.001)\Line(35.999,8.999)(28.001,1.001)
\end{picture}%
$ defined in $\left(  \ref{defvx}\right)  $, then these two diagrams become
the 1-loop diagrams for the $AAA$ amplitude in $\left(  \ref{triwti}\right)
$. Similarly, if we replace the circled cross $%
\begin{picture}(12,12) (10,-8)
\SetWidth{0.5}
\SetColor{Black}
\COval(16,-4)(5.657,5.657)(45.0){Black}{White}\Line
(13.172,-6.828)(18.828,-1.172)\Line(13.172,-1.172)(18.828,-6.828)
\end{picture}%
$ by the black dot $%
\begin{picture}(14,6) (-7,3)
\SetWidth{0.5}
\SetColor{Black}
\Vertex(0,6){3}
\end{picture}%
$ defined in $\left(  \ref{defvx}\right)  $, then the two diagrams become the
1-loop diagrams for the $AA\phi_{2}$ amplitude in $\left(  \ref{triwti}%
\right)  $. Thus the two diagrams in $\left(  \ref{a-01}\right)  $ represent
the left side of $\left(  \ref{triwti}\right)  $. Furthermore, according to
Appendix \ref{abcct}, only Levi-Civita tensor terms survive in the regularized
amplitudes for both $AAA$ and $AA\phi_{2}$. A 3-point 1PI function is linearly
divergent and the 2nd order term of its Taylor series expanded with respect to
its external momenta, henceforth called the $T_{2}$ term, is convergent. Thus
$T_{2}\left[  \Gamma_{F}^{\left(  1\right)  }\left(  A^{\mu},A^{\nu},\phi
_{2}\right)  \right]  $ is convergent and may be easily evaluated with any cut
point. The result
\begin{equation}
\lim_{n\rightarrow4}T_{2}\left[  \Gamma_{F}^{\left(  1\right)  }\left(
A^{\mu},A^{\nu},\phi_{2}\right)  \right]  =\frac{ig^{3}}{12\pi^{2}M}%
\epsilon^{\mu\nu\rho\sigma}k_{1\rho}k_{2\sigma} \label{fia21}%
\end{equation}
is unambiguously defined.

If the composite vertex $%
\begin{picture}(12,12) (10,-8)
\SetWidth{0.5}
\SetColor{Black}
\COval(16,-4)(5.657,5.657)(45.0){Black}{White}\Line
(13.172,-6.828)(18.828,-1.172)\Line(13.172,-1.172)(18.828,-6.828)
\end{picture}%
$ is detached from each of the two diagrams in $\left(  \ref{a-01}\right)  $,
both diagrams then become identical to%
\begin{equation}%
\begin{picture}(80,36) (41,-14)
\SetWidth{0.5}
\SetColor{Black}
\Line(60.001,0.001)(67.999,7.999)\Line(60.001,7.999)(67.999,0.001)
\Line(92.001,0.001)(99.999,7.999)\Line(92.001,7.999)(99.999,0.001)
\Arc
[arrow,arrowpos=0.5,arrowlength=2.5,arrowwidth=1,arrowinset=0.2](80,4)(16,-0,180)
\Arc
[arrow,arrowpos=0.5,arrowlength=2.5,arrowwidth=1,arrowinset=0.2](80,4)(16,-180,0)
\Text(56,-4)[]{\normalsize{\Black{$\nu$}}}
\Text(104,-4)[]{\normalsize{\Black{$\mu$}}}
\end{picture}
\label{a-01-1}%
\end{equation}
Since the component diagrams in $\left(  \ref{a-01}\right)  $ may be generated
by all possible insertions of the composite vertex $%
\begin{picture}(12,12) (10,-8)
\SetWidth{0.5}
\SetColor{Black}
\COval(16,-4)(5.657,5.657)(45.0){Black}{White}\Line
(13.172,-6.828)(18.828,-1.172)\Line(13.172,-1.172)(18.828,-6.828)
\end{picture}%
$ into the internal lines of $\left(  \ref{a-01-1}\right)  $, the diagram in
$\left(  \ref{a-01-1}\right)  $ will be called the generator for the Ward
identity $\left(  \ref{triwti}\right)  $.

By making a cut at the $\bar{\psi}-A^{\mu}-\psi$ vertex and then by repeated
use of $\left(  \ref{gbbwti}\right)  $ and $\left(  \ref{wti1}\right)  $, the
sum of the two diagrams in $\left(  \ref{a-01}\right)  $ becomes%
\begin{equation}%
\begin{picture}(306,32) (15,-22)
\SetWidth{0.5}
\SetColor{Black}
\Line(16,0)(64,0)
\Line(96,0)(144,0)
\Line(176,0)(224,0)
\Line(256,0)(304,0)
\Line
[arrow,arrowpos=1,arrowlength=2.5,arrowwidth=1,arrowinset=0.2](192,5)(178,5)
\Line
[arrow,arrowpos=1,arrowlength=2.5,arrowwidth=1,arrowinset=0.2](288,5)(298,5)
\COval(288,0)(5.657,5.657)(45.0){Black}{White}\Line
(285.172,-2.828)(290.828,2.828)\Line(285.172,2.828)(290.828,-2.828)
\Line(300.001,-3.999)(307.999,3.999)\Line(300.001,3.999)(307.999,-3.999)
\Line(268.001,-3.999)(275.999,3.999)\Line(268.001,3.999)(275.999,-3.999)
\Line(220.001,-3.999)(227.999,3.999)\Line(220.001,3.999)(227.999,-3.999)
\Line(204.001,-3.999)(211.999,3.999)\Line(204.001,3.999)(211.999,-3.999)
\COval(192,0)(5.657,5.657)(45.0){Black}{White}\Line
(189.172,-2.828)(194.828,2.828)\Line(189.172,2.828)(194.828,-2.828)
\Line(140.001,-3.999)(147.999,3.999)\Line(140.001,3.999)(147.999,-3.999)
\COval(128,0)(5.657,5.657)(45.0){Black}{White}\Line
(125.172,-2.828)(130.828,2.828)\Line(125.172,2.828)(130.828,-2.828)
\Line(108.001,-3.999)(115.999,3.999)\Line(108.001,3.999)(115.999,-3.999)
\Line(60.001,-3.999)(67.999,3.999)\Line(60.001,3.999)(67.999,-3.999)
\Line(44.001,-3.999)(51.999,3.999)\Line(44.001,3.999)(51.999,-3.999)
\COval(32,0)(5.657,5.657)(45.0){Black}{White}\Line
(29.172,-2.828)(34.828,2.828)\Line(29.172,2.828)(34.828,-2.828)
\Text(160,0)[]{\normalsize{\Black{$=$}}}
\Text(82,0)[]{\normalsize{\Black{$+$}}}
\Text(242,0)[]{\normalsize{\Black{$+$}}}
\Text(48,-16)[]{\normalsize{\Black{$\nu$}}}
\Text(112,-16)[]{\normalsize{\Black{$\nu$}}}
\Text(208,-16)[]{\normalsize{\Black{$\nu$}}}
\Text(272,-16)[]{\normalsize{\Black{$\nu$}}}
\Text(64,-16)[]{\normalsize{\Black{$\mu$}}}
\Text(144,-16)[]{\normalsize{\Black{$\mu$}}}
\Text(224,-16)[]{\normalsize{\Black{$\mu$}}}
\Text(304,-16)[]{\normalsize{\Black{$\mu$}}}
\end{picture}
\label{ewti2}%
\end{equation}
in which the horizontal line is supposed to be an open fermion line flowing to
the left. We emphasize that the identity $\left(  \ref{ewti2}\right)  $
remains satisfied when $n\neq4$ if we adopt the rightmost $\gamma_{5}$
dimensional regularization for every term in the identity. Calling the
momentum for the fermion line entering the cut point as $\ell$, we find that
these two amplitudes are, respectively,%
\[
\left(  -ig^{3}L\right)  \frac{1}{\not \ell +\not k  _{1}+\not k  _{2}%
-m}\gamma^{\nu}L\frac{1}{\not \ell +\not k  _{2}-m}\gamma^{\mu}L=L\hat
{M}\left(  \ell\right)  L
\]
and%
\[
\frac{1}{\not \ell -m}\gamma^{\nu}L\frac{1}{\not \ell +\not k  _{2}+\not k
_{3}-m}\left(  ig^{3}R\right)  \gamma^{\mu}L=-\hat{M}\left(  \ell
+k_{3}\right)  L,
\]
where $\hat{M}\left(  \ell\right)  $ stands for $-ig^{3}\frac{1}%
{\not \ell +\not k  _{1}+\not k  _{2}-m}\gamma^{\nu}L\frac{1}{\not \ell
+\not k  _{2}-m}\gamma^{\mu}$. Since the fermion lines form a closed loop, the
trace of the expressions above will be taken. With $\gamma_{5}$ rightmost
positioned, $Tr\left(  L\hat{M}\left(  \ell\right)  L\right)  $ may be reduced
to $Tr\left(  \hat{M}\left(  \ell\right)  L\right)  $ and the amplitudes
corresponding to the last two diagrams in $\left(  \ref{ewti2}\right)  $ are
related by a shift of the momentum variable. Since it is legitimate to shift
the loop momentum by a finite amount after regularization, the regularized
amplitude of $\left(  \ref{ewti2}\right)  $ vanishes after integration.

Note that the first two cut diagrams in $\left(  \ref{ewti2}\right)  $ may be
generated by attaching the composite vertex $%
\begin{picture}(12,12) (10,-8)
\SetWidth{0.5}
\SetColor{Black}
\COval(16,-4)(5.657,5.657)(45.0){Black}{White}\Line
(13.172,-6.828)(18.828,-1.172)\Line(13.172,-1.172)(18.828,-6.828)
\end{picture}%
$ in all possible manners consistent with Feynman rules to the cut diagram%
\begin{equation}%
\begin{picture}(50,27) (15,-27)
\SetWidth{0.5}
\SetColor{Black}
\Line(16,-5)(48,-5)
\Line(44.001,-8.999)(51.999,-1.001)\Line(44.001,-1.001)(51.999,-8.999)
\Line(28.001,-8.999)(35.999,-1.001)\Line(28.001,-1.001)(35.999,-8.999)
\Text(32,-21)[]{\normalsize{\Black{$\nu$}}}
\Text(48,-21)[]{\normalsize{\Black{$\mu$}}}
\end{picture}
\label{an-a9}%
\end{equation}
obtained by cutting the generator diagram $\left(  \ref{a-01-1}\right)  $ at
the $\bar{\psi}-A^{\mu}-\psi$ vertex. It is convenient to view the identity
that the regularized amplitude of $\left(  \ref{ewti2}\right)  $ vanishes as
being generated by the cut generator in $\left(  \ref{an-a9}\right)  $. To
summarize, if we choose the $\bar{\psi}-A^{\mu}-\psi$ vertex as the cut point
for the generator $\left(  \ref{a-01-1}\right)  $, construct the component
diagrams by attaching $%
\begin{picture}(12,12) (10,-8)
\SetWidth{0.5}
\SetColor{Black}
\COval(16,-4)(5.657,5.657)(45.0){Black}{White}\Line
(13.172,-6.828)(18.828,-1.172)\Line(13.172,-1.172)(18.828,-6.828)
\end{picture}%
$, anti-commute $\gamma_{5}$ to the rightmost position, and then dimensionally
regularize the coefficients in front of $\gamma_{5}$, the regularized
amplitudes so obtained satisfy the Ward identity $\left(  \ref{triwti}\right)
$.

Similarly, we may open up the fermion loop by choosing the $\bar{\psi}-A^{\nu
}-\psi$ vertex as the cut point, follow through the same arguments, and arrive
at another set of amplitudes for $\Gamma_{F}^{\left(  1\right)  }\left(
A^{\mu},A^{\nu},A^{\rho}\right)  $ and $\Gamma_{F}^{\left(  1\right)  }\left(
A^{\mu},A^{\nu},\phi_{2}\right)  $. Such amplitudes may also be obtained from
the interchange of $\left(  \mu,k_{1}\right)  \Leftrightarrow\left(  \nu
,k_{2}\right)  $ on the previously defined $\Gamma_{F}^{\left(  1\right)
}\left(  A^{\mu},A^{\nu},A^{\rho}\right)  $ and $\Gamma_{F}^{\left(  1\right)
}\left(  A^{\mu},A^{\nu},\phi_{2}\right)  $. Since $\epsilon^{\mu\nu\rho
\sigma}k_{1\rho}k_{2\sigma}$ is invariant under such exchange, the amplitude
$T_{2}\left[  \Gamma_{F}^{\left(  1\right)  }\left(  A^{\mu},A^{\nu},\phi
_{2}\right)  \right]  $ remains the same. Therefore, it may appear in order
that the result of $\left(  \ref{fia21}\right)  $ is consistent with the Ward
identity $\left(  \ref{triwti}\right)  $, $T_{1}\left[  \Gamma_{F}^{\left(
1\right)  }\left(  A^{\mu},A^{\nu},A^{\rho}\right)  \right]  $ (1st order term
in the Taylor series expansion) should be defined such that%
\begin{equation}
-ik_{3}^{\rho}\lim_{n\rightarrow4}T_{1}\left[  \Gamma_{F}^{\left(  1\right)
}\left(  A^{\mu},A^{\nu},A^{\rho}\right)  \right]  =\frac{ig^{3}}{12\pi^{2}%
}\epsilon^{\mu\nu\rho\sigma}k_{1\rho}k_{2\sigma}. \label{wtif3}%
\end{equation}

However, we will show, in the immediate following, that the above condition
$\left(  \ref{wtif3}\right)  $ for $AAA$ amplitude is inconsistent with the
Bose permutation symmetry. By definition, $T_{1}\left[  \Gamma_{F}^{\left(
1\right)  }\left(  A^{\mu},A^{\nu},A^{\rho}\right)  \right]  $ is a product of
a Levi-Civita tensor and a linear combination of the independent external
momenta $k_{1}$ and $k_{2}$. From relativistic covariance, we must have%
\begin{equation}
T_{1}\left[  \Gamma_{F}^{\left(  1\right)  }\left(  A^{\mu},A^{\nu},A^{\rho
};k_{1},k_{2},k_{3}\right)  \right]  =\epsilon^{\mu\nu\rho\sigma}\left(
C_{1}k_{1\sigma}+C_{2}k_{2\sigma}\right)  \label{bosesym1}%
\end{equation}
where $C_{1}$ and $C_{2}$ are dimensionless constants. The Bose symmetry under
the exchange of $\left(  A^{\mu},k_{1}\right)  \Leftrightarrow\left(  A^{\nu
},k_{2}\right)  $ gives
\[
T_{1}\left[  \Gamma_{F}^{\left(  1\right)  }\left(  A^{\mu},A^{\nu},A^{\rho
};k_{1},k_{2},k_{3}\right)  \right]  =T_{1}\left[  \Gamma_{F}^{\left(
1\right)  }\left(  A^{\nu},A^{\mu},A^{\rho};k_{2},k_{1},k_{3}\right)  \right]
,
\]
which is equivalent to%
\begin{equation}
\epsilon^{\mu\nu\rho\sigma}\left(  C_{1}k_{1\sigma}+C_{2}k_{2\sigma}\right)
=\epsilon^{\nu\mu\rho\sigma}\left(  C_{1}k_{2\sigma}+C_{2}k_{1\sigma}\right)
=\epsilon^{\mu\nu\rho\sigma}\left(  -C_{2}k_{1\sigma}-C_{1}k_{2\sigma}\right)
\label{cc1}%
\end{equation}
Similarly, the Bose symmetry for the exchange of $\left(  A^{\nu}%
,k_{2}\right)  \Leftrightarrow\left(  A^{\rho},k_{3}\right)  $ yields%
\[
T_{1}\left[  \Gamma_{F}^{\left(  1\right)  }\left(  A^{\mu},A^{\nu},A^{\rho
};k_{1},k_{2},k_{3}\right)  \right]  =T_{1}\left[  \Gamma_{F}^{\left(
1\right)  }\left(  A^{\mu},A^{\rho},A^{\nu};k_{1},k_{3},k_{2}\right)  \right]
\]
and
\begin{equation}
\epsilon^{\mu\nu\rho\sigma}\left(  C_{1}k_{1\sigma}+C_{2}k_{2\sigma}\right)
=\epsilon^{\mu\rho\nu\sigma}\left(  C_{1}k_{1\sigma}+C_{2}k_{3\sigma}\right)
=\epsilon^{\mu\nu\rho\sigma}\left(  \left(  C_{2}-C_{1}\right)  k_{1\sigma
}+C_{2}k_{2\sigma}\right)  . \label{cc2}%
\end{equation}
To meet $\left(  \ref{cc1}\right)  $ and $\left(  \ref{cc2}\right)  $, we must
have $C_{1}=C_{2}=0$. Consequently,%
\begin{equation}
T_{1}\left[  \Gamma_{F}^{\left(  1\right)  }\left(  A^{\mu},A^{\nu},A^{\rho
}\right)  \right]  =0. \label{t1aaa}%
\end{equation}
This result contradicts the result $\left(  \ref{wtif3}\right)  $ derived on
the basis of the validity of the Ward identity $\left(  \ref{triwti}\right)
$, showing that this Ward identity for the triangular diagrams is not
consistent with the Bose permutation symmetry.

Note that the diagrams used in the graphical identity $\left(  \ref{ewti2}%
\right)  $ with the $\bar{\psi}-A^{\mu}-\psi$ vertex as the cut point are not
Bose symmetric. Nor are those with the $\bar{\psi}-A^{\nu}-\psi$ vertex as the
cut point. Nor are the sum of these two sets of diagrams. This is because we
have left out the third vertex as a cut point. In fact, if we choose the
vertex $\bar{\psi}-A^{\rho}-\psi$ to be the cut point for $\Gamma_{F}^{\left(
1\right)  }\left(  A^{\mu},A^{\nu},A^{\rho}\right)  $ and the vertex
$\bar{\psi}-\phi_{2}-\psi$ to be the cut point for $\Gamma_{F}^{\left(
1\right)  }\left(  A^{\mu},A^{\nu},\phi_{2}\right)  $, the left side of
$\left(  \ref{triwti}\right)  $ is now diagrammatically expressed as%
\begin{equation}%
\begin{picture}(135,32) (15,-22)
\SetWidth{0.5}
\SetColor{Black}
\Line(16,0)(64,0)
\Line(96,0)(144,0)
\Line(124.001,-3.999)(131.999,3.999)\Line(124.001,3.999)(131.999,-3.999)
\COval(144,0)(5.657,5.657)(45.0){Black}{White}\Line
(141.172,-2.828)(146.828,2.828)\Line(141.172,2.828)(146.828,-2.828)
\Line(108.001,-3.999)(115.999,3.999)\Line(108.001,3.999)(115.999,-3.999)
\Line(44.001,-3.999)(51.999,3.999)\Line(44.001,3.999)(51.999,-3.999)
\Line(28.001,-3.999)(35.999,3.999)\Line(28.001,3.999)(35.999,-3.999)
\COval(64,0)(5.657,5.657)(45.0){Black}{White}\Line
(61.172,-2.828)(66.828,2.828)\Line(61.172,2.828)(66.828,-2.828)
\Text(82,0)[]{\normalsize{\Black{$+$}}}
\Text(32,-16)[]{\normalsize{\Black{$\nu$}}}
\Text(128,-16)[]{\normalsize{\Black{$\nu$}}}
\Text(48,-16)[]{\normalsize{\Black{$\mu$}}}
\Text(112,-16)[]{\normalsize{\Black{$\mu$}}}
\end{picture}
\label{wti3}%
\end{equation}
which, by making use of $\left(  \ref{gbbwti}\right)  $, is expanded into%
\begin{equation}%
\begin{picture}(302,32) (15,-22)
\SetWidth{0.5}
\SetColor{Black}
\Line(16,0)(64,0)
\Line(96,0)(144,0)
\Line(176,0)(224,0)
\Line(256,0)(304,0)
\Line
[arrow,arrowpos=1,arrowlength=2.5,arrowwidth=1,arrowinset=0.2](64,5)(54,5)
\Line
[arrow,arrowpos=1,arrowlength=2.5,arrowwidth=1,arrowinset=0.2](304,5)(314,5)
\Line(284.001,-3.999)(291.999,3.999)\Line(284.001,3.999)(291.999,-3.999)
\Line(268.001,-3.999)(275.999,3.999)\Line(268.001,3.999)(275.999,-3.999)
\Line(204.001,-3.999)(211.999,3.999)\Line(204.001,3.999)(211.999,-3.999)
\Line(188.001,-3.999)(195.999,3.999)\Line(188.001,3.999)(195.999,-3.999)
\Line(124.001,-3.999)(131.999,3.999)\Line(124.001,3.999)(131.999,-3.999)
\Line(108.001,-3.999)(115.999,3.999)\Line(108.001,3.999)(115.999,-3.999)
\Line(44.001,-3.999)(51.999,3.999)\Line(44.001,3.999)(51.999,-3.999)
\Line(28.001,-3.999)(35.999,3.999)\Line(28.001,3.999)(35.999,-3.999)
\Text(164,0)[]{\normalsize{\Black{$+$}}}
\Text(82,0)[]{\normalsize{\Black{$+$}}}
\Text(242,0)[]{\normalsize{\Black{$+$}}}
\Text(32,-16)[]{\normalsize{\Black{$\nu$}}}
\Text(112,-16)[]{\normalsize{\Black{$\nu$}}}
\Text(208,-16)[]{\normalsize{\Black{$\nu$}}}
\Text(288,-16)[]{\normalsize{\Black{$\nu$}}}
\Text(48,-16)[]{\normalsize{\Black{$\mu$}}}
\Text(128,-16)[]{\normalsize{\Black{$\mu$}}}
\Text(192,-16)[]{\normalsize{\Black{$\mu$}}}
\Text(272,-16)[]{\normalsize{\Black{$\mu$}}}
\Line
[arrow,arrowpos=1,arrowlength=2.5,arrowwidth=1,arrowinset=0.2](224,5)(214,5)
\Line
[arrow,arrowpos=1,arrowlength=2.5,arrowwidth=1,arrowinset=0.2](144,5)(154,5)
\COval(144,0)(5.657,5.657)(45.0){Black}{White}\Line
(141.172,-2.828)(146.828,2.828)\Line(141.172,2.828)(146.828,-2.828)
\COval(64,0)(5.657,5.657)(45.0){Black}{White}\Line
(61.172,-2.828)(66.828,2.828)\Line(61.172,2.828)(66.828,-2.828)
\COval(224,0)(5.657,5.657)(45.0){Black}{White}\Line
(221.172,-2.828)(226.828,2.828)\Line(221.172,2.828)(226.828,-2.828)
\COval(304,0)(5.657,5.657)(45.0){Black}{White}\Line
(301.172,-2.828)(306.828,2.828)\Line(301.172,2.828)(306.828,-2.828)
\end{picture}
\label{wti4}%
\end{equation}
Let $\ell$ be the momentum of the fermion line entering the cut point. The
symbol
{
\begin{picture}(20,12) (8,-8)
\SetWidth{0.5}
\SetColor{Black}
\LongArrow(15,2)(27,2)
\COval(15,-3)(5.66,5.66)(45.0){Black}{White}\Line
(12.17,-5.83)(17.83,-0.17)\Line(12.17,-0.17)(17.83,-5.83)
\end{picture}
}
on the right of either the second diagram or the fourth diagram in the above
figure is to be replaced by $gR\left(  \not \ell -m\right)  $ at the right-end
before moving $\gamma_{5}$\ to the rightmost position. If the factor of
$\left(  \not \ell -m\right)  $ at the right-end of the second (fourth)
diagram annihilates the fermion propagator $\frac{i}{\not \ell -m}$\ at the
left-end, the amplitude so obtained will cancel the amplitude of the third
(first) diagram. But in our scheme of dimensional regularization, we continue
to $n\neq4$ after positioning $\gamma_{5}$ at the rightmost site. This
rightmost $\gamma_{5}$ may stand between the $\frac{1}{\not \ell -m}$ at the
left-end and the $\left(  \not \ell -m\right)  $ at the right-end to prevent
their annihilation in the trace. In fact, when $n\neq4$, the symbol
{
\begin{picture}(20,12) (8,-8)
\SetWidth{0.5}
\SetColor{Black}
\LongArrow(15,2)(27,2)
\COval(15,-3)(5.66,5.66)(45.0){Black}{White}\Line
(12.17,-5.83)(17.83,-0.17)\Line(12.17,-0.17)(17.83,-5.83)
\end{picture}
}
should be replaced by the expression%
\begin{equation}
\left(  gR\left(  \not \ell -m\right)  \right)  |_{rightmost\text{ }\gamma
_{5}}=g\left(  \not \ell L-mR\right)  =gR\left(  \not \ell -m\right)
-g\not \ell _{\Delta}\gamma_{5}. \label{lot}%
\end{equation}
The last term $-g\not \ell _{\Delta}\gamma_{5}$ in the above is the leftover
after the cancellation. To evaluate the total amplitude of $\left(
\ref{wti4}\right)  $ by dimensional regularization, we only need to take into
account the contribution from the leftover terms. Furthermore, due to the
presence of $\not \ell _{\Delta}$, only the divergent orders in the Taylor
series expansion with respect to the external momenta may contribute to the
$n\rightarrow4$ limit. In particular, the leftover amplitude from the second
diagram of $\left(  \ref{wti4}\right)  $ is%
\begin{equation}
ig^{3}\int\frac{d^{n}\ell}{\left(  2\pi\right)  ^{n}}Tr\left(  \frac
{1}{\not \ell -m}\gamma^{\mu}\frac{\not \ell -\not k  _{1}}{\left(  \ell
-k_{1}\right)  ^{2}-m^{2}}\gamma^{\nu}\frac{\not \ell +\not k  _{3}}{\left(
\ell+k_{3}\right)  ^{2}-m^{2}}\not \ell _{\Delta}\gamma_{5}\right) \nonumber
\end{equation}
The evaluation of the above amplitude in the limit $n\rightarrow4$ is greatly
simplified by keeping only the $T_{2}$ order term to yield the result%
\begin{equation}
-\frac{ig^{3}}{8\pi^{2}}\epsilon^{\mu\nu\rho\sigma}k_{1\rho}k_{2\sigma}
\label{lt2}%
\end{equation}
In figure $\left(  \ref{wti4}\right)  $, the fourth diagram may be obtained
from the second diagram by the exchange $\left(  \mu,k_{1}\right)
\Leftrightarrow\left(  \nu,k_{2}\right)  $. Therefore, the leftover amplitude
due to the former diagram is the same as that due to the latter diagram and
the total amplitude of $\left(  \ref{wti3}\right)  $ in the limit
$n\rightarrow4$ is equal to twice the amount of $\left(  \ref{lt2}\right)  $,%
\begin{equation}
-\frac{ig^{3}}{4\pi^{2}}\epsilon^{\mu\nu\rho\sigma}k_{1\rho}k_{2\sigma},
\label{fw45}%
\end{equation}
which is also equal to $-M$ times thrice the amplitude of $T_{2}\left[
\Gamma_{F}^{\left(  1\right)  }\left(  A^{\mu},A^{\nu},\phi_{2}\right)
\right]  $ in $\left(  \ref{fia21}\right)  $. We have thus shown that the Ward
identity $\left(  \ref{triwti}\right)  $\ is not satisfied if the fermion
loops are cut open as in $\left(  \ref{wti3}\right)  $.

The values of $T_{2}\left[  -ik_{3}^{\rho}\Gamma_{F}^{\left(  1\right)
}\left(  A^{\mu},A^{\nu},A^{\rho}\right)  \right]  $ and $-MT_{2}\left[
\Gamma_{F}^{\left(  1\right)  }\left(  A^{\mu},A^{\nu},\phi_{2}\right)
\right]  $ evaluated with respect to the three possible cut points, after
factoring out $-ig^{3}\epsilon^{\mu\nu\rho\sigma}k_{1\rho}k_{2\sigma}$, are
summarized in the following table:
\begin{equation}%
\begin{tabular}
[c]{cccc}%
Cut point & $-ikAAA$ & $-MAA\phi_{2}$ & Sum\\
$%
\begin{picture}(12,12) (27,-1)
\SetWidth{0.5}
\SetColor{Black}
\Text(32,-2)[]{\normalsize{\Black{$\mu$}}}
\SetWidth{0.5}
\Line(28.001,8.999)(35.999,1.001)\Line(35.999,8.999)(28.001,1.001)
\end{picture}%
$ & $-\frac{1}{12\pi^{2}}$ & $\frac{1}{12\pi^{2}}$ & $0$\\
$%
\begin{picture}(12,12) (27,-1)
\SetWidth{0.5}
\SetColor{Black}
\Text(32,-2)[]{\normalsize{\Black{$\nu$}}}
\SetWidth{0.5}
\Line(28.001,8.999)(35.999,1.001)\Line(35.999,8.999)(28.001,1.001)
\end{picture}%
$ & $-\frac{1}{12\pi^{2}}$ & $\frac{1}{12\pi^{2}}$ & $0$\\
$%
\begin{picture}(12,12) (10,-8)
\SetWidth{0.5}
\SetColor{Black}
\COval(16,-4)(5.657,5.657)(45.0){Black}{White}\Line
(13.172,-6.828)(18.828,-1.172)\Line(13.172,-1.172)(18.828,-6.828)
\end{picture}%
$ & $\frac{1}{6\pi^{2}}$ & $\frac{1}{12\pi^{2}}$ & $\frac{1}{4\pi^{2}}$%
\end{tabular}
\label{tb2}%
\end{equation}

We have learned that $T_{1}\left[  \Gamma_{F}^{\left(  1\right)  }\left(
A^{\mu},A^{\nu},A^{\rho}\right)  \right]  $ vanishes if total permutation
symmetry is built into its component diagrams. The amplitude $\Gamma
_{F}^{\left(  1\right)  }\left(  A^{\mu},A^{\nu},A^{\rho}\right)  $ obtained
by averaging over the three amplitudes corresponding to the three different
cut points chosen at the vertices of $\bar{\psi}-A^{\mu}-\psi$, $\bar{\psi
}-A^{\nu}-\psi$, and $\bar{\psi}-A^{\rho}-\psi$ satisfies the permutation
symmetry. The amplitude $T_{1}\left[  \Gamma_{F}^{\left(  1\right)  }\left(
A^{\mu},A^{\nu},A^{\rho}\right)  \right]  $ so obtained therefore vanishes as
can be verified by summing over the three coefficients on the second column in
the above table $\left(  \ref{tb2}\right)  $. The amplitude $T_{2}\left[
\Gamma_{F}^{\left(  1\right)  }\left(  A^{\mu},A^{\nu},\phi_{2}\right)
\right]  $ is convergent and its value, which is\ independent of the cut point
chosen, remains equal to $\left(  \ref{fia21}\right)  $. If we take the
average of the $T_{2}$ order term for the left-hand side of the Ward identity
$\left(  \ref{triwti}\right)  $ over the three cuts, this average value does
not vanish but is equal to $-MT_{2}\left[  \Gamma_{F}^{\left(  1\right)
}\left(  A^{\mu},A^{\nu},\phi_{2}\right)  \right]  $. Since the permutation
symmetry of Bose statistics must be obeyed, we have to pay the price of losing
the validity of a Ward identity, and conclude that there exists an anomaly.

\subsection{Anomaly Compensating Fermion Field}

We have just observed that the ambiguity in choosing the cut point for the
1-loop $AAA$ amplitude owes its origin to the singular behavior of its
integrand. As a result, the Ward identity for this amplitude is not obeyed and
there is an anomaly. Let us add to the theory another fermion field
$\psi^{\prime}$ with a coupling constant $-g$ for $\psi_{L}^{\prime}$ equal to
the negative of the coupling constant $g$ for $\psi_{L}$. The covariant
derivative for $\psi_{L}^{\prime}$ is
\[
D_{\mu}\psi_{L}^{\prime}=\left(  \partial_{\mu}-igA_{\mu}\right)  \psi
_{L}^{\prime}%
\]
in contrast to the covariant derivative for $\psi_{L}$:%
\[
D_{\mu}\psi_{L}=\left(  \partial_{\mu}+igA_{\mu}\right)  \psi_{L}.
\]
The Lagrangian for such a theory is given by
\begin{equation}
L_{eff}^{\prime}=L_{eff}+\bar{\psi}_{L}^{\prime}\left(  i\not D  \right)
\psi_{L}^{\prime}+\bar{\psi}_{R}^{\prime}\left(  i\not \partial \right)
\psi_{R}^{\prime}-\sqrt{2}f^{\prime}\left(  \bar{\psi}_{L}^{\prime}%
\phi^{\dagger}\psi_{R}^{\prime}+\bar{\psi}_{R}^{\prime}\phi\psi_{L}^{\prime
}\right)  \label{Lanf}%
\end{equation}
where $L_{eff}$ is defined in $\left(  \ref{e2-3}\right)  $. Note the coupling
$f^{\prime}$ does not need to be the same as the $f$ in $\left(
\ref{e2-1}\right)  $ and the masses for the two fermion fields may not be
equal. The amplitude for a 1-loop $AAA$ diagram with the fermion loop due to
the $\psi^{\prime}$ field is proportional to $\left(  -g\right)  ^{3}$ and
cancels the logarithmically divergent term of the amplitude due to the $\psi$
field provided that we have synchronized the cut-point positions for both the
$\psi$ and $\psi^{\prime}$ fermion loops. The 1-loop $AAA$ amplitude is
convergent and cut point independent. Therefore the theory defined by $\left(
\ref{Lanf}\right)  $ is free of the 1-loop anomaly.

\section{Two-Loop Triangular Diagrams\label{2ltd}}

A straightforward calculation of the 2-loop anomaly is lengthy
\cite{AB,JL,MB,HC} without incorporating a gauge invariant regularization. By
utilizing the basic diagrammatic identities $\left(  \ref{gbbwti}\right)  $
and $\left(  \ref{wti1}\right)  $, we will be able to choose all rightmost
positions for the $\gamma_{5}$ to be in consistency with gauge invariance and
prove, without laborious calculation, the vanishing of the 2-loop anomaly. To
simplify the presentation in this section, we will only consider the subset of
2-loop triangular diagrams with one fermion loop and one internal vector meson
line. Other types of 2-loop triangular diagrams can be handled similarly
without additional difficulty and will be addressed in Appendix \ref{GWTI}.
For this restricted type of diagrams, the triangular Ward identity is the
identity that equates the sum of amplitudes for the following 12 diagrams to
zero.%
\begin{equation}%
\begin{picture}(372,111) (-5,-5)
\SetWidth{0.5}
\SetColor{Black}
\Line
[arrow,arrowpos=0.5,arrowlength=2.5,arrowwidth=1,arrowinset=0.2](150,101)(130,71)
\Line
[arrow,arrowpos=0.5,arrowlength=2.5,arrowwidth=1,arrowinset=0.2](130,71)(170,71)
\Line
[arrow,arrowpos=0.5,arrowlength=2.5,arrowwidth=1,arrowinset=0.2](170,71)(150,101)
\Line(167,68)(173,74)\Line(167,74)(173,68)
\Line(127,68)(133,74)\Line(127,74)(133,68)
\COval(150,101)(4.243,4.243)(45.0){Black}{White}\Line
(147.879,98.879)(152.121,103.121)\Line(147.879,103.121)(152.121,98.879)
\Line
[arrow,arrowpos=0.5,arrowlength=2.5,arrowwidth=1,arrowinset=0.2](30,101)(10,71)
\Line
[arrow,arrowpos=0.5,arrowlength=2.5,arrowwidth=1,arrowinset=0.2](10,71)(50,71)
\Line
[arrow,arrowpos=0.5,arrowlength=2.5,arrowwidth=1,arrowinset=0.2](50,71)(30,101)
\Line(47,68)(53,74)\Line(47,74)(53,68)
\Line(7,68)(13,74)\Line(7,74)(13,68)
\COval(30,101)(4.243,4.243)(45.0){Black}{White}\Line
(27.879,98.879)(32.121,103.121)\Line(27.879,103.121)(32.121,98.879)
\Line
[arrow,arrowpos=0.5,arrowlength=2.5,arrowwidth=1,arrowinset=0.2](90,101)(70,71)
\Line
[arrow,arrowpos=0.5,arrowlength=2.5,arrowwidth=1,arrowinset=0.2](70,71)(110,71)
\Line
[arrow,arrowpos=0.5,arrowlength=2.5,arrowwidth=1,arrowinset=0.2](110,71)(90,101)
\Line(107,68)(113,74)\Line(107,74)(113,68)
\Line(67,68)(73,74)\Line(67,74)(73,68)
\COval(90,101)(4.243,4.243)(45.0){Black}{White}\Line
(87.879,98.879)(92.121,103.121)\Line(87.879,103.121)(92.121,98.879)
\PhotonArc(21.209,86.093)(10.114,61.726,224.534){-1.5}{6.5}
\PhotonArc(90,71)(10,-180,0){-1.5}{6.5}
\PhotonArc(161,84)(9.434,-57.995,122.005){-1.5}{6.5}
\Line
[arrow,arrowpos=0.5,arrowlength=2.5,arrowwidth=1,arrowinset=0.2](329,101)(309,71)
\Line
[arrow,arrowpos=0.5,arrowlength=2.5,arrowwidth=1,arrowinset=0.2](309,71)(349,71)
\Line
[arrow,arrowpos=0.5,arrowlength=2.5,arrowwidth=1,arrowinset=0.2](349,71)(329,101)
\Line(346,68)(352,74)\Line(346,74)(352,68)
\Line(306,68)(312,74)\Line(306,74)(312,68)
\COval(329,101)(4.243,4.243)(45.0){Black}{White}\Line
(326.879,98.879)(331.121,103.121)\Line(326.879,103.121)(331.121,98.879)
\Line
[arrow,arrowpos=0.5,arrowlength=2.5,arrowwidth=1,arrowinset=0.2](269,101)(249,71)
\Line
[arrow,arrowpos=0.5,arrowlength=2.5,arrowwidth=1,arrowinset=0.2](249,71)(289,71)
\Line
[arrow,arrowpos=0.5,arrowlength=2.5,arrowwidth=1,arrowinset=0.2](289,71)(269,101)
\Line(286,68)(292,74)\Line(286,74)(292,68)
\Line(246,68)(252,74)\Line(246,74)(252,68)
\COval(269,101)(4.243,4.243)(45.0){Black}{White}\Line
(266.879,98.879)(271.121,103.121)\Line(266.879,103.121)(271.121,98.879)
\Line
[arrow,arrowpos=0.5,arrowlength=2.5,arrowwidth=1,arrowinset=0.2](209,101)(189,71)
\Line
[arrow,arrowpos=0.5,arrowlength=2.5,arrowwidth=1,arrowinset=0.2](189,71)(229,71)
\Line
[arrow,arrowpos=0.5,arrowlength=2.5,arrowwidth=1,arrowinset=0.2](229,71)(209,101)
\Line(226,68)(232,74)\Line(226,74)(232,68)
\Line(186,68)(192,74)\Line(186,74)(192,68)
\COval(209,101)(4.243,4.243)(45.0){Black}{White}\Line
(206.879,98.879)(211.121,103.121)\Line(206.879,103.121)(211.121,98.879)
\Photon(256,82)(264,71){-1}{4}
\Photon(202,90)(217,90){-1}{4}
\Photon(344,80)(337,71){-1}{4}
\Line
[arrow,arrowpos=0.5,arrowlength=2.5,arrowwidth=1,arrowinset=0.2](150,41)(130,11)
\Line
[arrow,arrowpos=0.5,arrowlength=2.5,arrowwidth=1,arrowinset=0.2](130,11)(170,11)
\Line
[arrow,arrowpos=0.5,arrowlength=2.5,arrowwidth=1,arrowinset=0.2](170,11)(150,41)
\Line(167,8)(173,14)\Line(167,14)(173,8)
\Line(127,8)(133,14)\Line(127,14)(133,8)
\COval(150,41)(4.243,4.243)(45.0){Black}{White}\Line
(147.879,38.879)(152.121,43.121)\Line(147.879,43.121)(152.121,38.879)
\Line
[arrow,arrowpos=0.5,arrowlength=2.5,arrowwidth=1,arrowinset=0.2](30,41)(10,11)
\Line
[arrow,arrowpos=0.5,arrowlength=2.5,arrowwidth=1,arrowinset=0.2](10,11)(50,11)
\Line
[arrow,arrowpos=0.5,arrowlength=2.5,arrowwidth=1,arrowinset=0.2](50,11)(30,41)
\Line(47,8)(53,14)\Line(47,14)(53,8)
\Line(7,8)(13,14)\Line(7,14)(13,8)
\COval(30,41)(4.243,4.243)(45.0){Black}{White}\Line
(27.879,38.879)(32.121,43.121)\Line(27.879,43.121)(32.121,38.879)
\Line
[arrow,arrowpos=0.5,arrowlength=2.5,arrowwidth=1,arrowinset=0.2](90,41)(70,11)
\Line
[arrow,arrowpos=0.5,arrowlength=2.5,arrowwidth=1,arrowinset=0.2](70,11)(110,11)
\Line
[arrow,arrowpos=0.5,arrowlength=2.5,arrowwidth=1,arrowinset=0.2](110,11)(90,41)
\Line(107,8)(113,14)\Line(107,14)(113,8)
\Line(67,8)(73,14)\Line(67,14)(73,8)
\COval(90,41)(4.243,4.243)(45.0){Black}{White}\Line
(87.879,38.879)(92.121,43.121)\Line(87.879,43.121)(92.121,38.879)
\PhotonArc(21.209,26.093)(10.114,61.726,224.534){-1.5}{6.5}
\PhotonArc(90,11)(10,-180,0){-1.5}{6.5}
\PhotonArc(161,24)(9.434,-57.995,122.005){-1.5}{6.5}
\Line
[arrow,arrowpos=0.5,arrowlength=2.5,arrowwidth=1,arrowinset=0.2](329,41)(309,11)
\Line
[arrow,arrowpos=0.5,arrowlength=2.5,arrowwidth=1,arrowinset=0.2](309,11)(349,11)
\Line
[arrow,arrowpos=0.5,arrowlength=2.5,arrowwidth=1,arrowinset=0.2](349,11)(329,41)
\Line(346,8)(352,14)\Line(346,14)(352,8)
\Line(306,8)(312,14)\Line(306,14)(312,8)
\COval(329,41)(4.243,4.243)(45.0){Black}{White}\Line
(326.879,38.879)(331.121,43.121)\Line(326.879,43.121)(331.121,38.879)
\Line
[arrow,arrowpos=0.5,arrowlength=2.5,arrowwidth=1,arrowinset=0.2](269,41)(249,11)
\Line
[arrow,arrowpos=0.5,arrowlength=2.5,arrowwidth=1,arrowinset=0.2](249,11)(289,11)
\Line
[arrow,arrowpos=0.5,arrowlength=2.5,arrowwidth=1,arrowinset=0.2](289,11)(269,41)
\Line(286,8)(292,14)\Line(286,14)(292,8)
\Line(246,8)(252,14)\Line(246,14)(252,8)
\COval(269,41)(4.243,4.243)(45.0){Black}{White}\Line
(266.879,38.879)(271.121,43.121)\Line(266.879,43.121)(271.121,38.879)
\Line
[arrow,arrowpos=0.5,arrowlength=2.5,arrowwidth=1,arrowinset=0.2](209,41)(189,11)
\Line
[arrow,arrowpos=0.5,arrowlength=2.5,arrowwidth=1,arrowinset=0.2](189,11)(229,11)
\Line
[arrow,arrowpos=0.5,arrowlength=2.5,arrowwidth=1,arrowinset=0.2](229,11)(209,41)
\Line(226,8)(232,14)\Line(226,14)(232,8)
\Line(186,8)(192,14)\Line(186,14)(192,8)
\COval(209,41)(4.243,4.243)(45.0){Black}{White}\Line
(206.879,38.879)(211.121,43.121)\Line(206.879,43.121)(211.121,38.879)
\Photon(256,22)(264,11){-1}{4}
\Photon(202,30)(217,30){-1}{4}
\Photon(344,20)(337,11){-1}{4}
\Text(230,1)[]{\normalsize{\Black{$\mu$}}}
\Text(10,61)[]{\normalsize{\Black{$\mu$}}}
\Text(70,61)[]{\normalsize{\Black{$\mu$}}}
\Text(130,61)[]{\normalsize{\Black{$\mu$}}}
\Text(190,61)[]{\normalsize{\Black{$\mu$}}}
\Text(250,61)[]{\normalsize{\Black{$\mu$}}}
\Text(310,61)[]{\normalsize{\Black{$\mu$}}}
\Text(350,1)[]{\normalsize{\Black{$\mu$}}}
\Text(290,1)[]{\normalsize{\Black{$\mu$}}}
\Text(50,1)[]{\normalsize{\Black{$\mu$}}}
\Text(110,1)[]{\normalsize{\Black{$\mu$}}}
\Text(170,1)[]{\normalsize{\Black{$\mu$}}}
\Text(10,1)[]{\normalsize{\Black{$\nu$}}}
\Text(50,61)[]{\normalsize{\Black{$\nu$}}}
\Text(110,61)[]{\normalsize{\Black{$\nu$}}}
\Text(170,61)[]{\normalsize{\Black{$\nu$}}}
\Text(230,61)[]{\normalsize{\Black{$\nu$}}}
\Text(290,61)[]{\normalsize{\Black{$\nu$}}}
\Text(350,61)[]{\normalsize{\Black{$\nu$}}}
\Text(310,1)[]{\normalsize{\Black{$\nu$}}}
\Text(250,1)[]{\normalsize{\Black{$\nu$}}}
\Text(190,1)[]{\normalsize{\Black{$\nu$}}}
\Text(130,1)[]{\normalsize{\Black{$\nu$}}}
\Text(70,1)[]{\normalsize{\Black{$\nu$}}}
\end{picture}
\label{a-02}%
\end{equation}

In the above figure, the fermion loop is the arrowed loop and the wavy lines
are vector meson lines. The above 12 diagrams are also the ones generated by
attaching the composite $%
\begin{picture}(12,12) (10,-8)
\SetWidth{0.5}
\SetColor{Black}
\COval(16,-4)(5.657,5.657)(45.0){Black}{White}\Line
(13.172,-6.828)(18.828,-1.172)\Line(13.172,-1.172)(18.828,-6.828)
\end{picture}%
$ vertex in all possible manners consistent with Feynman rules to the
following three generator diagrams:
\begin{equation}%
\begin{picture}(240,36) (9,-30)
\SetWidth{0.5}
\SetColor{Black}
\Photon(208,4)(208,-28){-2}{6}
\Line(189,-15)(195,-9)\Line(189,-9)(195,-15)
\Line(221,-15)(227,-9)\Line(221,-9)(227,-15)
\Text(232,-20)[]{\normalsize{\Black{$\nu$}}}
\Line(109,-15)(115,-9)\Line(109,-9)(115,-15)
\Line(141,-15)(147,-9)\Line(141,-9)(147,-15)
\Text(104,-20)[]{\normalsize{\Black{$\nu$}}}
\Text(152,-20)[]{\normalsize{\Black{$\mu$}}}
\PhotonArc[clock](128,21.5)(25.5,-61.928,-118.072){-1.5}{5.5}
\Arc
[arrow,arrowpos=0.5,arrowlength=2.5,arrowwidth=1,arrowinset=0.2](128,-12)(16,-0,180)
\Arc
[arrow,arrowpos=0.5,arrowlength=2.5,arrowwidth=1,arrowinset=0.2](128,-12)(16,-180,0)
\Line(29,-15)(35,-9)\Line(29,-9)(35,-15)
\Line(61,-15)(67,-9)\Line(61,-9)(67,-15)
\Text(24,-20)[]{\normalsize{\Black{$\mu$}}}
\Text(72,-20)[]{\normalsize{\Black{$\nu$}}}
\PhotonArc[clock](48,21.5)(25.5,-61.928,-118.072){-1.5}{5.5}
\Arc
[arrow,arrowpos=0.5,arrowlength=2.5,arrowwidth=1,arrowinset=0.2](48,-12)(16,-0,180)
\Arc
[arrow,arrowpos=0.5,arrowlength=2.5,arrowwidth=1,arrowinset=0.2](48,-12)(16,-180,0)
\Arc
[arrow,arrowpos=0.75,arrowlength=2.5,arrowwidth=1,arrowinset=0.2](208,-12)(16,-90,90)
\Arc
[arrow,arrowpos=0.75,arrowlength=2.5,arrowwidth=1,arrowinset=0.2](208,-12)(16,90,270)
\Text(184,-20)[]{\normalsize{\Black{$\mu$}}}
\end{picture}
\label{a-02-1a}%
\end{equation}

In order not to give an asymmetric treatment to any of the external fields, we
will refrain from using cut points at the vertices connecting to external
fields. A cut point is deemed illegitimate if it is positioned at a vertex
connecting to an external field line. For the diagrams in $\left(
\ref{a-02}\right)  $, cut points at $%
\begin{picture}(12,12) (27,-1)
\SetWidth{0.5}
\SetColor{Black}
\Text(32,-2)[]{\normalsize{\Black{$\mu$}}}
\SetWidth{0.5}
\Line(28.001,8.999)(35.999,1.001)\Line(35.999,8.999)(28.001,1.001)
\end{picture}%
$, $%
\begin{picture}(12,12) (27,-1)
\SetWidth{0.5}
\SetColor{Black}
\Text(32,-2)[]{\normalsize{\Black{$\nu$}}}
\SetWidth{0.5}
\Line(28.001,8.999)(35.999,1.001)\Line(35.999,8.999)(28.001,1.001)
\end{picture}%
$ or $%
\begin{picture}(12,12) (10,-8)
\SetWidth{0.5}
\SetColor{Black}
\COval(16,-4)(5.657,5.657)(45.0){Black}{White}\Line
(13.172,-6.828)(18.828,-1.172)\Line(13.172,-1.172)(18.828,-6.828)
\end{picture}%
$ vertices are illegitimate.

Because we do not position $\gamma_{5}$ inside a self-energy or
vertex-correction sub-diagram on an open fermion line in our prescription, it
is also appropriate to avoid cutting the fermion loops at such positions. A
position for $\gamma_{5}$ will be called proper if it is not located within a
divergent 1PI sub-diagram such as a self-energy insertion or a vertex
correction. For a fermion loop, a cut and the corresponding cut point will be
called proper if the cut is not made within a divergent self-energy insertion
or vertex correction sub-diagram.

It will be shown in Appendix \ref{abcct} that we only need to use cut points
at the endpoints of fermion lines to evaluate the Levi-Civita tensor terms.
Therefore, for each of the above 12 diagrams in $\left(  \ref{a-02}\right)  $,
we may choose the cut point to be the one and the only one that is proper,
legitimate and located at the end of an internal fermion line. For
convenience, the sum of regularized amplitudes for the 12 cut diagrams so
obtained will be denoted by $S_{12}$. Since none of the external fields is
given a preferential treatment, the $AAA$ amplitude obtained by replacing the
composite vertex $%
\begin{picture}(12,12) (10,-8)
\SetWidth{0.5}
\SetColor{Black}
\COval(16,-4)(5.657,5.657)(45.0){Black}{White}\Line
(13.172,-6.828)(18.828,-1.172)\Line(13.172,-1.172)(18.828,-6.828)
\end{picture}%
$ with the vertex $%
\begin{picture}(12,12) (27,-1)
\SetWidth{0.5}
\SetColor{Black}
\Text(32,-2)[]{\normalsize{\Black{$\rho$}}}
\SetWidth{0.5}
\Line(28.001,8.999)(35.999,1.001)\Line(35.999,8.999)(28.001,1.001)
\end{picture}%
$ in each of the 12 component diagrams in $S_{12}$ is symmetric with respect
to the permutation of the three external vector fields $A^{\mu}$, $A^{\nu}$
and $A^{\rho}$. This permutation symmetry ensures that the $T_{1}$ order term
of the $AAA$ amplitude vanishes and therefore $S_{12}$ is superficially convergent.

There are many cancellations among the amplitudes for the 12 component
diagrams in $S_{12}$. For example,\ making the cut at the endpoint of the
fermion line connecting to the 1-loop fermion self-energy insertion on the
second diagram in $\left(  \ref{a-02-1a}\right)  $ yields the cut generator%
\begin{equation}%
\begin{picture}(57,28) (5,-8)
\SetWidth{0.5}
\SetColor{Black}
\Line(60,8)(10,8)
\PhotonArc(50,8)(10,-0,180){-1.5}{7.5}
\Line(27,5)(33,11)\Line(27,11)(33,5)
\Text(30,-2)[]{\normalsize{\Black{$\nu$}}}
\Line(17,5)(23,11)\Line(17,11)(23,5)
\Text(20,-2)[]{\normalsize{\Black{$\mu$}}}
\end{picture}
\label{a-02-4a}%
\end{equation}
which gives, after attaching $%
\begin{picture}(12,12) (10,-8)
\SetWidth{0.5}
\SetColor{Black}
\COval(16,-4)(5.657,5.657)(45.0){Black}{White}\Line
(13.172,-6.828)(18.828,-1.172)\Line(13.172,-1.172)(18.828,-6.828)
\end{picture}%
$ in all possible manners that are consistent with Feynman rules, the
following four component diagrams:%
\begin{equation}%
\begin{picture}(293,28) (9,-8)
\SetWidth{0.5}
\SetColor{Black}
\Text(40,-2)[]{\normalsize{\Black{$\nu$}}}
\Text(30,-2)[]{\normalsize{\Black{$\mu$}}}
\SetWidth{0.5}
\Line(70,8)(10,8)
\COval(20,8)(4.243,4.243)(45.0){Black}{White}\Line
(17.879,5.879)(22.121,10.121)\Line(17.879,10.121)(22.121,5.879)
\Line(27,5)(33,11)\Line(27,11)(33,5)
\Line(37,5)(43,11)\Line(37,11)(43,5)
\PhotonArc(60,8)(10,-0,180){-1.5}{7.5}
\Line(150,8)(90,8)
\COval(110,8)(4.243,4.243)(45.0){Black}{White}\Line
(107.879,5.879)(112.121,10.121)\Line(107.879,10.121)(112.121,5.879)
\Line(97,5)(103,11)\Line(97,11)(103,5)
\Line(117,5)(123,11)\Line(117,11)(123,5)
\PhotonArc(140,8)(10,-0,180){-1.5}{7.5}
\Line(230,8)(170,8)
\COval(200,8)(4.243,4.243)(45.0){Black}{White}\Line
(197.879,5.879)(202.121,10.121)\Line(197.879,10.121)(202.121,5.879)
\Line(177,5)(183,11)\Line(177,11)(183,5)
\Line(187,5)(193,11)\Line(187,11)(193,5)
\PhotonArc(220,8)(10,-0,180){-1.5}{7.5}
\Line(300,8)(250,8)
\COval(290,8)(4.243,4.243)(45.0){Black}{White}\Line
(287.879,5.879)(292.121,10.121)\Line(287.879,10.121)(292.121,5.879)
\Line(257,5)(263,11)\Line(257,11)(263,5)
\Line(267,5)(273,11)\Line(267,11)(273,5)
\PhotonArc(290,8)(10,-0,180){-1.5}{7.5}
\Text(80,8)[]{\normalsize{\Black{$+$}}}
\Text(160,8)[]{\normalsize{\Black{$+$}}}
\Text(240,8)[]{\normalsize{\Black{$+$}}}
\Text(100,-2)[]{\normalsize{\Black{$\mu$}}}
\Text(180,-2)[]{\normalsize{\Black{$\mu$}}}
\Text(260,-2)[]{\normalsize{\Black{$\mu$}}}
\Text(120,-2)[]{\normalsize{\Black{$\nu$}}}
\Text(190,-2)[]{\normalsize{\Black{$\nu$}}}
\Text(270,-2)[]{\normalsize{\Black{$\nu$}}}
\end{picture}
\label{a-03}%
\end{equation}
Using $\left(  \ref{gbbwti}\right)  $ and $\left(  \ref{wti1}\right)  $
repeatedly, the above expression can be reduced to%
\begin{equation}%
\begin{picture}(146,28) (9,-8)
\SetWidth{0.5}
\SetColor{Black}
\Line
[arrow,arrowpos=1,arrowlength=2.5,arrowwidth=1,arrowinset=0.2](20,12)(12,12)
\Text(40,-2)[]{\normalsize{\Black{$\nu$}}}
\Text(30,-2)[]{\normalsize{\Black{$\mu$}}}
\Line(70,8)(10,8)
\Line(27,5)(33,11)\Line(27,11)(33,5)
\Line(37,5)(43,11)\Line(37,11)(43,5)
\PhotonArc(60,8)(10,-0,180){-1.5}{7.5}
\Line(150,8)(90,8)
\Line(97,5)(103,11)\Line(97,11)(103,5)
\Line(107,5)(113,11)\Line(107,11)(113,5)
\PhotonArc(130,8)(10,-0,180){-1.5}{7.5}
\Text(80,8)[]{\normalsize{\Black{$-$}}}
\Text(100,-2)[]{\normalsize{\Black{$\mu$}}}
\Text(110,-2)[]{\normalsize{\Black{$\nu$}}}
\Line
[arrow,arrowpos=1,arrowlength=2.5,arrowwidth=1,arrowinset=0.2](150,12)(142,12)
\COval(150,8)(4.243,4.243)(45.0){Black}{White}\Line
(147.879,5.879)(152.121,10.121)\Line(147.879,10.121)(152.121,5.879)
\COval(20,8)(4.243,4.243)(45.0){Black}{White}\Line
(17.879,5.879)(22.121,10.121)\Line(17.879,10.121)(22.121,5.879)
\end{picture}
\label{a-04}%
\end{equation}
If we identify $f\left(  \ell_{1},\ell_{2}\right)  $ as the Feynman integrand
for the last diagram, where $\ell_{2}$ is the momentum of the vector meson
line and $\ell_{1}$ is the momentum of the leftmost fermion line, the Feynman
integrand corresponding to $\left(  \ref{a-04}\right)  $ is the difference of
two terms related by a shift of the loop momentum $\ell_{1}$. Specifically,
this sum is%
\begin{equation}
f\left(  \ell_{1}-k_{3},\ell_{2}\right)  -f\left(  \ell_{1},\ell_{2}\right)
\label{wcdif}%
\end{equation}
which vanishes upon carrying out the integration $\int d^{n}\ell_{1}d^{n}%
\ell_{2}$ under our scheme of rightmost $\gamma_{5}$ dimensional
regularization. The sum of amplitudes for the 4 diagrams in $\left(
\ref{a-03}\right)  $ therefore vanishes. Likewise, the sum of the 4 diagrams
obtained from $\left(  \ref{a-03}\right)  $ by making the exchange $\left(
\mu,k_{1}\right)  \Leftrightarrow\left(  \nu,k_{2}\right)  $ also vanishes. By
deleting those 4+4 diagrams from the 12 diagrams of $S_{12}$, we are left with
4 diagrams, each of which has a 1-loop vertex-correction sub-diagram. Let us
define $S_{4}$ to be the sum of amplitudes for these 4 remaining diagrams.
$S_{4}$ is equal to $S_{12}$ and is superficially convergent as well.

Unlike the first two diagrams in $\left(  \ref{a-02-1a}\right)  $, no proper
cut point is available for the third diagram. Making the cut at the endpoint
of the fermion line connecting to the 1-loop radiative correction for the
vertex $\bar{\psi}-A^{\nu}-\psi$ on the third diagram in $\left(
\ref{a-02-1a}\right)  $, we get the cut generator
\begin{equation}%
\begin{picture}(52,28) (5,-8)
\SetWidth{0.5}
\SetColor{Black}
\Text(40,-2)[]{\normalsize{\Black{$\nu$}}}
\Text(20,-2)[]{\normalsize{\Black{$\mu$}}}
\SetWidth{0.5}
\Line(50,8)(10,8)
\Line(17,5)(23,11)\Line(17,11)(23,5)
\Line(37,5)(43,11)\Line(37,11)(43,5)
\PhotonArc(40,8)(10,-0,180){-1.5}{7.5}
\end{picture}
\label{a-02-4b}%
\end{equation}
that, after attaching $%
\begin{picture}(12,12) (10,-8)
\SetWidth{0.5}
\SetColor{Black}
\COval(16,-4)(5.657,5.657)(45.0){Black}{White}\Line
(13.172,-6.828)(18.828,-1.172)\Line(13.172,-1.172)(18.828,-6.828)
\end{picture}%
$, yields the identity%
\begin{equation}%
\begin{picture}(298,28) (9,-8)
\SetWidth{0.5}
\SetColor{Black}
\Text(50,-2)[]{\normalsize{\Black{$\nu$}}}
\Text(30,-2)[]{\normalsize{\Black{$\mu$}}}
\SetWidth{0.5}
\Line(60,8)(10,8)
\COval(20,8)(4.243,4.243)(45.0){Black}{White}\Line
(17.879,5.879)(22.121,10.121)\Line(17.879,10.121)(22.121,5.879)
\Line(27,5)(33,11)\Line(27,11)(33,5)
\Line(47,5)(53,11)\Line(47,11)(53,5)
\PhotonArc(50,8)(10,-0,180){-1.5}{7.5}
\Line(130,8)(80,8)
\COval(100,8)(4.243,4.243)(45.0){Black}{White}\Line
(97.879,5.879)(102.121,10.121)\Line(97.879,10.121)(102.121,5.879)
\Line(87,5)(93,11)\Line(87,11)(93,5)
\Line(117,5)(123,11)\Line(117,11)(123,5)
\PhotonArc(120,8)(10,-0,180){-1.5}{7.5}
\Text(70,8)[]{\normalsize{\Black{$+$}}}
\Text(210,8)[]{\normalsize{\Black{$+$}}}
\Text(90,-2)[]{\normalsize{\Black{$\mu$}}}
\Text(120,-2)[]{\normalsize{\Black{$\nu$}}}
\Line(200,8)(150,8)
\COval(180,8)(4.243,4.243)(45.0){Black}{White}\Line
(177.879,5.879)(182.121,10.121)\Line(177.879,10.121)(182.121,5.879)
\Line(157,5)(163,11)\Line(157,11)(163,5)
\Line(187,5)(193,11)\Line(187,11)(193,5)
\PhotonArc(185,1.75)(16.25,22.62,157.38){-1.5}{9.5}
\Text(140,8)[]{\normalsize{\Black{$+$}}}
\Text(160,-2)[]{\normalsize{\Black{$\mu$}}}
\Text(190,-2)[]{\normalsize{\Black{$\nu$}}}
\Line(270,8)(220,8)
\COval(260,8)(4.243,4.243)(45.0){Black}{White}\Line
(257.879,5.879)(262.121,10.121)\Line(257.879,10.121)(262.121,5.879)
\Line(227,5)(233,11)\Line(227,11)(233,5)
\Line(247,5)(253,11)\Line(247,11)(253,5)
\PhotonArc(255,1.75)(16.25,22.62,157.38){-1.5}{9.5}
\Text(230,-2)[]{\normalsize{\Black{$\mu$}}}
\Text(250,-2)[]{\normalsize{\Black{$\nu$}}}
\Text(290,8)[]{\normalsize{\Black{$= \ 0$}}}
\end{picture}
\label{a-12}%
\end{equation}

For the first two diagrams in the above, the cut points are proper. But for
each of the last two diagrams, if we reconnect the beginning point and the
endpoint of the open fermion line to restore the original fermion loop, we see
that there is a sub-diagram of radiative correction for the vertex $\bar{\psi
}-A^{\mu}-\psi$. The cut point, being the endpoint of the fermion line in this
vertex correction sub-diagram, is improper. From here on in this section, we
will identity $S_{2}$ as the sum of the last two diagrams in $\left(
\ref{a-12}\right)  $. The sub-diagram of radiative correction for the vertex
$\bar{\psi}-A^{\mu}-\psi$ in each diagram of $S_{2}$ will be denoted by $H$.

For both diagrams in $S_{2}$, if the improper cut point inside $H$ is moved
out of $H$ to the endpoint of the fermion line connecting to $H$, the
relocated cut becomes proper and $S_{2}$ becomes%
\begin{equation}%
\begin{picture}(128,28) (9,-8)
\SetWidth{0.5}
\SetColor{Black}
\Text(30,-2)[]{\normalsize{\Black{$\nu$}}}
\Text(50,-2)[]{\normalsize{\Black{$\mu$}}}
\SetWidth{0.5}
\Line(60,8)(10,8)
\COval(20,8)(4.243,4.243)(45.0){Black}{White}\Line
(17.879,5.879)(22.121,10.121)\Line(17.879,10.121)(22.121,5.879)
\Line(27,5)(33,11)\Line(27,11)(33,5)
\Line(47,5)(53,11)\Line(47,11)(53,5)
\PhotonArc(50,8)(10,-0,180){-1.5}{7.5}
\Line(130,8)(80,8)
\COval(100,8)(4.243,4.243)(45.0){Black}{White}\Line
(97.879,5.879)(102.121,10.121)\Line(97.879,10.121)(102.121,5.879)
\Line(87,5)(93,11)\Line(87,11)(93,5)
\Line(117,5)(123,11)\Line(117,11)(123,5)
\PhotonArc(120,8)(10,-0,180){-1.5}{7.5}
\Text(70,8)[]{\normalsize{\Black{$+$}}}
\Text(120,-2)[]{\normalsize{\Black{$\mu$}}}
\Text(90,-2)[]{\normalsize{\Black{$\nu$}}}
\end{picture}
\label{a-05}%
\end{equation}
Since all the fermion lines and vertex factors sandwiched between the original
cut points in $S_{2}$ and the relocated ones in $\left(  \ref{a-05}\right)  $
lie within the sub-diagram $H$, the difference between $S_{2}$ and $\left(
\ref{a-05}\right)  $ may be expressed as a combination of terms with
$\gamma_{\Delta}$ factors stemming from the matrix product in $H$. These
$\gamma_{\Delta}$ factors may not be ignored\ if they are multiplied by pole
terms arising from the logarithmically divergent loop integrations due to the
sub-diagram $H$ or the overall diagram.

Making repetitive use of $\left(  \ref{gbbwti}\right)  $ and $\left(
\ref{wti1}\right)  $, $S_{2}$ can be transformed into%
\begin{equation}%
\begin{picture}(126,28) (5,-8)
\SetWidth{0.5}
\SetColor{Black}
\Text(70,8)[]{\normalsize{\Black{$+$}}}
\SetWidth{0.5}
\Line(60,8)(10,8)
\Line(17,5)(23,11)\Line(17,11)(23,5)
\Line(47,5)(53,11)\Line(47,11)(53,5)
\PhotonArc(45,1.75)(16.25,22.62,157.38){-1.5}{9.5}
\Text(20,-2)[]{\normalsize{\Black{$\mu$}}}
\Text(50,-2)[]{\normalsize{\Black{$\nu$}}}
\Line(130,8)(80,8)
\Line(87,5)(93,11)\Line(87,11)(93,5)
\Line(107,5)(113,11)\Line(107,11)(113,5)
\Text(90,-2)[]{\normalsize{\Black{$\mu$}}}
\Text(110,-2)[]{\normalsize{\Black{$\nu$}}}
\PhotonArc(115,1.75)(16.25,22.62,157.38){-1.5}{9.5}
\Line
[arrow,arrowpos=1,arrowlength=2.5,arrowwidth=1,arrowinset=0.2](120,12)(128,12)
\COval(120,8)(4.243,4.243)(45.0){Black}{White}\Line
(117.879,5.879)(122.121,10.121)\Line(117.879,10.121)(122.121,5.879)
\Line
[arrow,arrowpos=1,arrowlength=2.5,arrowwidth=1,arrowinset=0.2](40,12)(32,12)
\COval(40,8)(4.243,4.243)(45.0){Black}{White}\Line
(37.879,5.879)(42.121,10.121)\Line(37.879,10.121)(42.121,5.879)
\end{picture}
\label{a-07}%
\end{equation}
There is a similar transformation for $\left(  \ref{a-05}\right)  $. The
divergent loop integration of the sub-diagram $H$ only occurs in the $T_{0}$
term of $H$, denoted by $T_{0}\left[  H\right]  $, which is the amplitude of
$H$ with all the external momenta relative to $H$ set to zero. If we
substitute $T_{0}\left[  H\right]  $ for $H$ in either diagram of $\left(
\ref{a-07}\right)  $, the resulting amplitude must vanish because it depends
only one external momentum $k_{1}$ and two external polarizations $\mu$ and
$\nu$, which are insufficient to form a Levi-Civita tensor term. Thus the
divergent loop integral of $H$ does not contribute to $S_{2}$. Similarly, the
divergence of $H$ does not contribute to $\left(  \ref{a-05}\right)  $.

Let us define $\bar{S}_{2}$ to be the first two diagrams in $\left(
\ref{a-12}\right)  $. The two diagrams of $\left(  \ref{a-05}\right)  $ may be
obtained from $\bar{S}_{2}$ by making the interchange $\left(  \mu
,k_{1}\right)  \Leftrightarrow\left(  \nu,k_{2}\right)  $. Since $S_{4}%
$\ consists of the two diagrams of $\bar{S}_{2}$ and the two diagrams of
$\left(  \ref{a-05}\right)  $, it is symmetric under $\left(  \mu
,k_{1}\right)  \Leftrightarrow\left(  \nu,k_{2}\right)  $. In the Taylor
series expansion with respect to the external momenta for $S_{2}$, $\bar
{S}_{2}$ or $S_{4}$, only the second order $T_{2}$ term may have superficially
divergent Levi-Civita tensor terms. In addition, any such $T_{2}$ order term
is equal to some constant times $\epsilon^{\mu\nu\rho\sigma}k_{1\rho
}k_{2\sigma}$ which is invariant under $\left(  \mu,k_{1}\right)
\Leftrightarrow\left(  \nu,k_{2}\right)  $. Thus $T_{2}\left[  S_{4}\right]  $
is equal to twice $T_{2}\left[  \bar{S}_{2}\right]  $. Since we have shown
that $S_{4}$ is superficially convergent, $\bar{S}_{2}$ is also superficially
convergent. Furthermore, the identity $\left(  \ref{a-12}\right)  $, which is
equivalent to $\bar{S}_{2}+S_{2}=0$, ensures that $S_{2}$ is superficially
convergent as well. There is no divergent pole term to prevent the difference
between $S_{2}$ and $\left(  \ref{a-05}\right)  $ from vanishing in the limit
$n\rightarrow4$. As a consequence, both $S_{4}$ and $S_{12}$\ vanish in the
limit $n\rightarrow4$ and we have succeeded in regularizing and preserving the
2-loop triangular Ward identity under dimensional regularization.

In the 1-loop case, one of the three external vertices must be used as the cut
point and we have shown it is impossible to construct a set of diagrams to
satisfy both the triangular Ward identity and Bose permutation symmetry. For
the two-loop diagrams we have discussed here, there is the additional freedom
of choosing cut points at vertices connecting to internal vector meson lines.
As a result, we are able to construct diagrams that satisfy both the
triangular Ward identity and Bose permutation symmetry.

The $AA\phi_{2}\ $function is superficially convergent and its renormalized
amplitude can be calculated with any convenient choice of proper cut point.
Since the 2-loop triangular Ward identity can be regularized and renormalized
by minimal subtractions without violating Bose permutation symmetry, the
$T_{2}$ term of the renormalized $AA\phi_{2}$ amplitude can be expressed as a
linear combination of the $T_{1}$ term of the renormalized $AAA$ amplitude.
Knowing that $T_{1}\left[  AAA\right]  $ vanishes on the sole account of
permutation symmetry, $T_{2}\left[  AA\phi_{2}\right]  $ must vanish as well.
This condition has been verified by direct calculation \cite{HC} without using
dimensional regularization.

\section{Conclusion}

In this paper, we have found a simple and natural way to treat $\gamma_{5}$ in
dimensional regularization: moving all $\gamma_{5}$ matrices to the rightmost
position before analytically continuing the dimension. For amplitudes
corresponding diagrams without fermion loops, the amplitudes obtained with our
prescription automatically satisfy the Ward identities without further ado.

The rightmost position on a fermion loop is not defined. For this reason, we
introduce the concept of a cut point. We have found that the choice of a cut
point often conflicts with gauge invariance. From this vantage point, this
lack of a rightmost position is what breaks the Ward identities, leading to
triangular anomalies.

Applying our prescriptions to 1-loop triangular amplitudes, we reproduce
correctly the value of the triangular anomaly, verifying that our prescription
is applicable to diagrams with anomalies. For a 1-loop fermion self-energy
diagram or a 1-loop vertex correction diagram, positioning $\gamma_{5}$ within
the divergent 1PI diagram gives an amplitude differing from the amplitude
obtained with rightmost $\gamma_{5}$ by a finite amount, even after
subtraction of pole terms. Thus for a 2-loop diagram with a fermion loop and
with a 1-loop self-energy insertion or a 1-loop radiative vertex insertion, we
do not assign a point inside a divergent 1-loop sub-diagram as a cut point.
Furthermore, in order not to give a preferential role to any of the external
lines, we do not choose the point of the vertex connecting to an external
field line as a cut point. We have shown that this prescription of utilizing
proper and legitimate cut points enables us to regulate amplitudes in a gauge
invariant manner.%

\appendix\appendixpage

\section{Charge Conjugation Transformation\label{abcct}}

If we disregard terms involving fermion fields, the effective Lagrangian
$\left(  \ref{e2-3}\right)  $ is invariant under the following charge
conjugation transformation:%
\begin{align}
H  &  \rightarrow H\label{cctran}\\
\phi_{2}  &  \rightarrow-\phi_{2},A^{\mu}\rightarrow-A^{\mu},\bar
{c}\rightarrow-\bar{c},c\rightarrow-c.\nonumber
\end{align}
Fields that have odd (even) charge parity shorthanded as $C$-parity under this
transformation are classified as $C$-odd ($C$-even) fields. For non-fermion
fields, $H$ is $C$-even and non-$H$ fields are $C$-odd. We define the
$C$-parity of a Feynman diagram to be the product of the $C$-parities of its
non-fermion external lines. A vertex without fermion lines attached is always
$C$-even, so is a\ fermionless Feynman diagram. It is therefore impossible to
construct a $C$-odd diagram without including fermion lines or loops. For a
theory that does not involve $\gamma_{5}$, such as QED, the charge conjugation
transformation is a symmetry of its Lagrangian. A consequence of this symmetry
is the Furry theorem which states that any amplitude for an odd number of
external vector fields such as the $AAA$ amplitude vanishes in QED.

In four dimensional space, the charge conjugation transformation
$\psi\rightarrow C\bar{\psi}^{T}$ for fermion fields is effected by the matrix%
\begin{equation}
C=i\gamma^{2}\gamma^{0} \label{cmatrix}%
\end{equation}
that satisfies%
\begin{equation}
C\gamma^{\mu}C^{-1}=-\left(  \gamma^{\mu}\right)  ^{T}. \label{ccgt}%
\end{equation}
The above identity is based on the property that $\gamma^{0}$ and $\gamma^{2}$
are symmetric matrices while $\gamma^{1}$ and $\gamma^{3}$ are antisymmetric
in four dimensional space. It is not guaranteed that this property specific to
$n=4$ may be dimensionally continued such that $\left(  \ref{ccgt}\right)  $
holds when $n\neq4$.

We shall not assume the validity of $\left(  \ref{ccgt}\right)  $ when
$n\neq4$ and define instead the charge conjugation for a matrix product of $N$
$\gamma$ matrices $\hat{M}=\gamma^{\mu_{1}}\gamma^{\mu_{2}}..\gamma^{\mu_{N}}$
as $\hat{M}^{C}=\left(  -\gamma^{\mu_{N}}\right)  ..\left(  -\gamma^{\mu_{2}%
}\right)  \left(  -\gamma^{\mu_{1}}\right)  $ which is the product of the
negative of these $N$ $\gamma$ matrices in reversed order. When $n=4$, we may
make use of $\left(  \ref{ccgt}\right)  $ to verify straightforwardly that the
trace of $\hat{M}$ is the same as that of $\hat{M}^{C}$ or%
\begin{equation}
Tr\left(  \gamma^{\mu_{1}}\gamma^{\mu_{2}}..\gamma^{\mu_{N}}\right)
=Tr\left(  \left(  -\gamma^{\mu_{N}}\right)  ..\left(  -\gamma^{\mu_{2}%
}\right)  \left(  -\gamma^{\mu_{1}}\right)  \right)  \label{trcgt}%
\end{equation}
Since both sides in $\left(  \ref{trcgt}\right)  $ consist of terms that are
product of $g^{\mu_{\iota}\mu_{j}}$ metric tensors, the polarizations $\mu
_{1},\mu_{2},..\mu_{N}$ may be dimensionally continued beyond the first 4
dimensions so that $\left(  \ref{trcgt}\right)  $ is also valid when $n\neq4$.
The validity of%
\[
Tr\left(  \gamma^{\mu_{1}}\gamma^{\mu_{2}}..\gamma^{\mu_{N}}\gamma^{0}%
\gamma^{1}\gamma^{2}\gamma^{3}\right)  =Tr\left(  \gamma^{3}\gamma^{2}%
\gamma^{1}\gamma^{0}\left(  -\gamma^{\mu_{N}}\right)  ..\left(  -\gamma
^{\mu_{2}}\right)  \left(  -\gamma^{\mu_{1}}\right)  \right)
\]
also yields%
\begin{equation}
Tr\left(  \gamma^{\mu_{1}}\gamma^{\mu_{2}}..\gamma^{\mu_{N}}\gamma_{5}\right)
=Tr\left(  \gamma_{5}\left(  -\gamma^{\mu_{N}}\right)  ..\left(  -\gamma
^{\mu_{2}}\right)  \left(  -\gamma^{\mu_{1}}\right)  \right)  \label{trcgt5}%
\end{equation}
where $\mu_{1},\mu_{2},..\mu_{N}$ are allowed to be polarizations in arbitrary
$n$ dimensional space.

We will use the notation $\hat{M}_{DR}$ with the sub-index $_{DR}$ to indicate
that $\hat{M}_{DR}$ is the matrix product obtained from $\hat{M}$ by
anti-commuting all the $\gamma_{5}$ matrices with $\gamma$ matrices to the
right and then continuing to $n\neq4$. Conditions $\left(  \ref{trcgt}\right)
$ and $\left(  \ref{trcgt5}\right)  $ in the above may be summarized as%
\begin{equation}
Tr\left(  \hat{M}_{DR}\right)  =Tr\left(  \left(  \hat{M}_{DR}\right)
^{C}\right)  . \label{trdr}%
\end{equation}
The accumulated sign change resulting from moving a $\gamma_{5}$ in $\hat{M}$
to the rightmost position is equal to that from moving the corresponding
$\gamma_{5}$ in $\hat{M}^{C}$ to the leftmost position. We thus have
\begin{equation}
\left(  \hat{M}_{DR}\right)  ^{C}=\left(  \hat{M}^{C}\right)  _{DL}
\label{drdlc}%
\end{equation}
where the subscript $_{DL}$ means that the analytical continuation to $n\neq4$
starts from the expression obtained after anti-commuting all the $\gamma_{5}$
factors to the leftmost position. If the count of $\gamma^{\mu}$ matrices with
$\mu\in\left\{  0,1,2,3\right\}  $ in a matrix product is odd, the trace of
the matrix product is zero and so is its continuation from any form. Hence,
the $\gamma_{5}$ factor at the leftmost position of $\left(  \hat{M}%
^{C}\right)  _{DL}$ in a trace may be moved to the rightmost position to yield%
\begin{equation}
Tr\left(  \left(  \hat{M}^{C}\right)  _{DL}\right)  =Tr\left(  \left(  \hat
{M}^{C}\right)  _{DR}\right)  \label{trl2r}%
\end{equation}
Combining $\left(  \ref{trdr}\right)  $, $\left(  \ref{drdlc}\right)  $ and
$\left(  \ref{trl2r}\right)  $ in the above, we get%
\begin{equation}
Tr\left(  \hat{M}_{DR}\right)  =Tr\left(  \left(  \hat{M}^{C}\right)
_{DR}\right)  \label{trdr1}%
\end{equation}

Let $G$ be a Feynman diagram with a fermion loop that has been cut open at the
point $P$. The conjugate diagram $G^{C}$ is defined to be the diagram obtained
by reversing the direction of the fermion loop in $G$. The point $P$ remains
to be the cut point of $G^{C}$. If the cut point $P$ on $G$ is the endpoint of
a certain fermion line on the loop, it becomes the beginning point of the
reversed fermion line in $G^{C}$, and vice versa.

The identity $\left(  \ref{trdr1}\right)  $ may be utilized to show that
dimensionally regularized amplitudes for $G$ and $G^{C}$ are related. To be
more specific, let $F$ be a fermion loop attached by fields in the sequence
$\varphi_{1},\varphi_{2},...\varphi_{n}$ with inward momenta $k_{1}%
,k_{2},..k_{n}$ and the cut point is chosen to be the endpoint of the internal
fermion line flowing into the vertex of $\varphi_{1}$. The Feynman integrand
$I\left(  F\right)  $ for $F$ may be written as%
\begin{equation}
I\left(  F\right)  =Tr\left(
\begin{array}
[c]{c}%
\frac{i}{\not \ell -m}\varpi\left(  \varphi_{n}\right)  \frac{i}%
{\not \ell +\not k  _{1}+\not k  _{2}..+\not k  _{n-1}-m}\varpi\left(
\varphi_{n-1}\right)  ...\\
\times\frac{i}{\not \ell +\not k  _{1}+\not k  _{2}-m}\varpi\left(
\varphi_{2}\right)  \frac{i}{\not \ell +\not k  _{1}-m}\varpi\left(
\varphi_{1}\right)
\end{array}
\right)  _{DR} \label{cpef}%
\end{equation}
where $\ell$ is the loop momentum variable and the vertex factors are
\[
\varpi\left(  A^{\mu}\right)  =-igR\gamma^{\mu}L\text{, }\varpi\left(
\phi_{2}\right)  =f\gamma_{5}\text{ and }\varpi\left(  H\right)  =-if\text{.}%
\]
According to the identity $\left(  \ref{trdr1}\right)  $, performing the
charge conjugation operation on the matrix product inside the trace of
$\left(  \ref{cpef}\right)  $ leaves the value of $I\left(  F\right)  $
unchanged. Thus,%
\[
I\left(  F\right)  =Tr\left(
\begin{array}
[c]{c}%
\tilde{\varpi}\left(  \varphi_{1}\right)  \frac{i}{-\left(  \not \ell +\not k
_{1}\right)  -m}\tilde{\varpi}\left(  \varphi_{2}\right)  \frac{i}{-\left(
\not \ell +\not k  _{1}+\not k  _{2}\right)  -m}...\\
\times\tilde{\varpi}\left(  \varphi_{n-1}\right)  \frac{i}{-\left(
\not \ell +\not k  _{1}+\not k  _{2}..+\not k  _{n-1}\right)  -m}\tilde
{\varpi}\left(  \varphi_{n}\right)  \frac{i}{-\not \ell -m}%
\end{array}
\right)  _{DR}%
\]
where%
\[
\tilde{\varpi}\left(  A^{\mu}\right)  =igL\gamma^{\mu}R\text{, }\tilde{\varpi
}\left(  \phi_{2}\right)  =f\gamma_{5}\text{ and }\tilde{\varpi}\left(
H\right)  =-if\text{.}%
\]
We are allowed to make the transformation $\ell\rightarrow-\ell$ in carrying
out the $\int d^{n}\ell$ loop integration and arrive at%
\begin{equation}
\int d^{n}\ell I\left(  F\right)  =\int d^{n}\ell Tr\left(
\begin{array}
[c]{c}%
\tilde{\varpi}\left(  \varphi_{1}\right)  \frac{i}{\not \ell -\not k  _{1}%
-m}\tilde{\varpi}\left(  \varphi_{2}\right)  \frac{i}{\not \ell -\not k
_{1}-\not k  _{2}-m}...\\
\times\tilde{\varpi}\left(  \varphi_{n-1}\right)  \frac{i}{\not \ell -\not k
_{1}-\not k  _{2}..-\not k  _{n-1}-m}\tilde{\varpi}\left(  \varphi_{n}\right)
\frac{i}{\not \ell -m}%
\end{array}
\right)  _{DR} \label{cpef1}%
\end{equation}
On the other hand, the conjugate diagram $F^{C}$ is the fermion loop with the
external fields attached on the loop in the order of $\varphi_{n}%
,\varphi_{n-1},...\varphi_{1}$, and with the cut point being the beginning
point of the fermion line that leaves the vertex of $\varphi_{1}$. The Feynman
integrand for $F^{C}$ may be written as%
\begin{equation}
I\left(  F^{C}\right)  =Tr\left(
\begin{array}
[c]{c}%
\varpi\left(  \varphi_{1}\right)  \frac{i}{\not \ell -\not k  _{1}-m}%
\varpi\left(  \varphi_{2}\right)  \frac{i}{\not \ell -\not k  _{1}-\not k
_{2}-m}...\\
\times\varpi\left(  \varphi_{n-1}\right)  \frac{i}{\not \ell -\not k
_{1}-\not k  _{2}..-\not k  _{n-1}-m}\varpi\left(  \varphi_{n}\right)
\frac{i}{\not \ell -m}%
\end{array}
\right)  _{DR} \label{cpef2}%
\end{equation}
Let us observe that
\begin{align*}
\tilde{\varpi}\left(  A^{\mu}\right)   &  =-\varpi\left(  A^{\mu}\right)
|_{\gamma_{5}\rightarrow-\gamma_{5}},\\
\tilde{\varpi}\left(  \phi_{2}\right)   &  =-\varpi\left(  \phi_{2}\right)
|_{\gamma_{5}\rightarrow-\gamma_{5}},\\
\tilde{\varpi}\left(  H\right)   &  =\varpi\left(  H\right)  |_{\gamma
_{5}\rightarrow-\gamma_{5}}.
\end{align*}
These relationships demonstrate that if we insert an additional negative sign
in front of every $\gamma_{5}$, the vertex factors $\tilde{\varpi}\left(
A^{\mu}\right)  $, $\tilde{\varpi}\left(  \phi_{2}\right)  $ and
$\tilde{\varpi}\left(  H\right)  $ become $-\varpi\left(  A^{\mu}\right)  $,
$-\varpi\left(  \phi_{2}\right)  $ and $\varpi\left(  H\right)  $
respectively. Note also that the integrand in $\left(  \ref{cpef1}\right)  $
becomes the integrand $I\left(  F^{C}\right)  $ in $\left(  \ref{cpef2}%
\right)  $ if all the vertex factors $\tilde{\varpi}\left(  \varphi\right)  $
in $\left(  \ref{cpef1}\right)  $ are replaced by $\varpi\left(
\varphi\right)  $. Thus we have%
\begin{equation}
\int d^{n}\ell I\left(  F^{C}\right)  =\left(  -1\right)  ^{N_{C}\left(
F\right)  }\int d^{n}\ell I\left(  F\right)  |_{\gamma_{5}\rightarrow
-\gamma_{5}} \label{cei0}%
\end{equation}
where $N_{C}\left(  F\right)  $ is the number of $C$-odd fields in $\left\{
\varphi_{1},\varphi_{2},...\varphi_{n}\right\}  $ and $\left(  -1\right)
^{N_{C}\left(  F\right)  }$ is the $C$-parity of the diagram $F$ or $F^{C}$.
Decomposing the identity $\left(  \ref{cei0}\right)  $ into the $\gamma_{5}%
$-even part and the $\gamma_{5}$-odd part, we get%
\begin{equation}
\gamma_{5}\text{-even part of}\int d^{n}\ell I\left(  F^{C}\right)
=\gamma_{5}\text{-even part of }\left(  -1\right)  ^{N_{C}\left(  F\right)
}\int d^{n}\ell I\left(  F\right)  \label{cei1}%
\end{equation}
and%
\begin{equation}
\gamma_{5}\text{-odd part of}\int d^{n}\ell I\left(  F^{C}\right)  =\gamma
_{5}\text{-odd part of }\left(  -1\right)  ^{N_{C}\left(  F\right)  +1}\int
d^{n}\ell I\left(  F\right)  \text{.} \label{cei2}%
\end{equation}
If the fermion loop $F$ is a sub-diagram of a larger diagram $G$ that contains
no other fermion lines than those in $F$, then the $C$-parity of $G$ is equal
to the $C$-parity of $F$. Since the Feynman integrand for the complement of
$F$ in $G$ is the same as that for the complement of $F^{C}$ in $G^{C}$,
$\left(  \ref{cei0}\right)  $-$\left(  \ref{cei2}\right)  $ are also valid if
we replace $F$ with $G$. If $G$ is $C$-even, the $\gamma_{5}$-even part of the
dimensionally regularized amplitude of $G$ is equal to the $\gamma_{5}$-even
part of $G^{C}$ but the $\gamma_{5}$-odd part of $G$ is the negative of the
$\gamma_{5}$-odd part of $G^{C}$. If $G$ is $C$-odd, the $\gamma_{5}$-odd part
of $G$ is equal to the $\gamma_{5}$-odd part of $G^{C}$ and the $\gamma_{5}%
$-even part of $G$ is the negative of the $\gamma_{5}$-even part of $G^{C}$.

We will require that if a diagram $G$ is included as a component diagram, the
conjugate diagram $G^{C}$ must also be included (with, of course, suitable
adjustment of weighting factors). Since the $\gamma_{5}$-odd parts are
cancelled between $G$ and $G^{C}$ when $G$ is $C$-even, no Levi-Civita tensor
term is possible for $C$-even functions.

For $C$-odd functions, the $\gamma_{5}$-even parts are cancelled between $G$
and $G^{C}$. If we discard the $\gamma_{5}$-even part, either $G$ or $G^{C}$
suffices for the evaluation of the $C$-odd function. We will use the one whose
cut point is located at the endpoint of an internal fermion line. In other
words, the Levi-Civita tensor terms for $C$-odd functions may be evaluated by
diagrams whose cut points are restricted to the subset of endpoints of
internal fermion lines on the fermion loops.

\section{Green Functions and Ward Identities\label{GWTI}}

In the main context, we only consider Feynman diagrams in which the
non-fermion internal lines are the vector meson lines. To handle other types
of diagrams, we will make use of Green functions.

The Green function $G$ is the vacuum expectation value of a time-ordered
product. Specifically,
\begin{equation}
G\left(  O_{1}\left(  x_{1}\right)  ,O_{2}\left(  x_{2}\right)  ,...O_{n}%
\left(  x_{n}\right)  \right)  =T\left\langle O_{1}\left(  x_{1}\right)
O_{2}\left(  x_{2}\right)  ...O_{n}\left(  x_{n}\right)  \right\rangle ,
\label{gtdef}%
\end{equation}
where the operator $O_{i}\left(  x_{i}\right)  $ is either a field operator or
a product of field operators at the same space-time point $x_{i}$. The
connected Green function, denoted by $G_{c}$, is%
\[
G_{c}\left(  ...\right)  =all\,connected\,diagrams\,of\,G\left(  ...\right)
\]

We need a notation to indicate that some external lines of a Green function
are amputated. To denote a truncated external line, we underline the
corresponding field variable in the Green function. $i.e$.,
\begin{align}
G\left(  ...,\varphi_{i},...\right)   &  =D\left(  \varphi_{i},\varphi
_{j}\right)  G\left(  ...,\underline{\varphi_{j}},...\right) \label{udtp}\\
G_{c}\left(  ...,\varphi_{i},...\right)   &  =D\left(  \varphi_{i},\varphi
_{j}\right)  G_{c}\left(  ...,\underline{\varphi_{j}},...\right)  ,\nonumber
\end{align}
where the propagator $D\left(  \varphi_{i},\varphi_{j}\right)  $ is also the
two-point Green function,
\[
D\left(  \varphi_{i},\varphi_{j}\right)  =G\left(  \varphi_{i},\varphi
_{j}\right)  .
\]
Note that in $\left(  \ref{udtp}\right)  $ the space-time dependence of the
field variable $\varphi_{i}$ is lumped into the index $i$ and the Einstein
summation convention for the repeated index $j$\ is extended to include
summation over all possible field types and integration of space-time points.
The fully truncated Green function $\Gamma$ is the connected Green function
with all field variables underlined.
\[
\Gamma\left(  \varphi_{1},\varphi_{2},...,\varphi_{n}\right)  =G_{c}\left(
\underline{\varphi_{1},\varphi_{2},...,\varphi_{n}}\right)
\]
In particular, $\Gamma\left(  \varphi_{i},\varphi_{j}\right)  $ is the inverse
propagator.
\[
\Gamma\left(  \varphi_{i},\varphi_{j}\right)  =G\left(  \underline{\varphi
_{i},\varphi_{j}}\right)  =D^{-1}\left(  \varphi_{j},\varphi_{i}\right)
\]
For a composite operator $\hat{O}$, which is a product of field operators at
the same space-time point, we define%
\[
\Gamma\left(  \varphi_{1},\varphi_{2},...,\varphi_{n},\hat{O}\right)
=G_{c}\left(  \underline{\varphi_{1},\varphi_{2},...,\varphi_{n}},\hat
{O}\right)  \text{.}%
\]
Note that to avoid misinterpretations, $\hat{O}$ is forbidden to be a single
field operator in the above identification. The tree order part of a\ Green
function $\digamma$, which may be any of the above $G,$ $G_{c}$ or $\Gamma$
function, will be denoted by the notation $\digamma^{\left(  0\right)  }$ with
the superscript $\left(  0\right)  $. The Fourier transform of a\ Green
function $\digamma$ is labeled by an additional group of momentum variables
and is related to its counterpart in the coordinate space by%
\begin{align*}
&  \digamma\left(  \varphi_{1}\left(  x_{1}\right)  ,\varphi_{2}\left(
x_{2}\right)  ,...,\varphi_{n}\left(  x_{n}\right)  \right) \\
&  =\int\frac{dk_{1}}{\left(  2\pi\right)  ^{4}}\frac{dk_{2}}{\left(
2\pi\right)  ^{4}}..\frac{dk_{n-1}}{\left(  2\pi\right)  ^{4}}e^{-i\left(
k_{1}x_{1}+k_{2}x_{2}+...k_{n}x_{n}\right)  }\digamma\left(  \varphi
_{1},\varphi_{2},...,\varphi_{n};k_{1},k_{2}...k_{n}\right)  ,
\end{align*}
where $k_{1}+k_{2}+...+k_{n}=0$. We will omit the momentum variables
$k_{1},k_{2}...k_{n}$ for the Fourier transform if there is little chance of confusion.

\subsection{Basic Graphical Identities}

The BRST invariance leads to a number of Ward identities which form an
important part of the foundation on which renormalizability is based. These
identities can be formally derived in the following way. The vacuum state
$|0>$ in the theory satisfies%
\begin{equation}
Q|0>=0\label{brsqvac}%
\end{equation}
where $Q$ is the BRST charge. The commutator (anticommutator) of $iQ$ with a
non-ghost (ghost) field is equal to the BRST variation of the field. Because
of $\left(  \ref{brsqvac}\right)  $, we have%
\[
T\left\langle 0\left\vert iQ\varphi_{1}\left(  x_{1}\right)  \varphi
_{2}\left(  x_{2}\right)  ...\right\vert 0\right\rangle =0
\]
where $\varphi_{i}$ is a field operator. By moving $iQ$ to the right until it
operates on $|0>$ and vanishes, we get%
\begin{align}
&  T\left\langle 0\left\vert \delta\left(  \varphi_{1}\left(  x_{1}\right)
\varphi_{2}\left(  x_{2}\right)  ...\right)  \right\vert 0\right\rangle
\label{bidt}\\
&  =T\left\langle 0\left\vert \delta\varphi_{1}\left(  x_{1}\right)
\varphi_{2}\left(  x_{2}\right)  ...\right\vert 0\right\rangle \pm
\ T\left\langle 0\left\vert \varphi_{1}\left(  x_{1}\right)  \delta\varphi
_{2}\left(  x_{2}\right)  ...\right\vert 0\right\rangle \pm\ ...=0\nonumber
\end{align}
The relative sign between terms is determined by the positions of the ghost
fields. The above BRST identity is formal and its renormalized version may not
be satisfied when anomaly exists. But the tree order terms are finite and
always satisfy the BRST identity provided the Lagrangian is BRST invariant. To
facilitate the discussions for higher loop order terms, we will introduce
graphical notations for some basic tree order identities.

The BRST variation for any field variable $\varphi\left(  x\right)  $ in
general may be decomposed as
\begin{equation}
\delta\varphi\left(  x\right)  =\delta_{1}\varphi\left(  x\right)  +\delta
_{2}\varphi\left(  x\right)  ,\label{dcbrsv}%
\end{equation}
in which $\delta_{1}\varphi\left(  x\right)  $ is a linear superposition of
field variables and $\delta_{2}\varphi\left(  x\right)  $ is a product of the
ghost field $c\left(  x\right)  $ and another field variable at the same
space-time point $x$. For the Abelian-Higgs theory with the Lagrangian
$\left(  \ref{e2-3}\right)  $, non-vanishing $\delta_{1}\varphi$ are
$\delta_{1}A^{\mu}=\partial^{\mu}c$ and $\delta_{1}\phi_{2}=-Mc$, and
non-vanishing $\delta_{2}\varphi$ are $\delta_{2}H=gc\phi_{2}$, $\delta
_{2}\phi_{2}=-gcH$ and $\delta_{2}\psi_{L}=-igc\psi_{L}$.

By $\left(  \ref{bidt}\right)  $, we have%
\begin{equation}
T\left\langle 0\left\vert \left(  \delta\bar{c}\left(  z\right)  \right)
\varphi_{i}\right\vert 0\right\rangle _{\left(  0\right)  }=T\left\langle
0\left\vert \bar{c}\left(  z\right)  \delta\varphi_{i}\right\vert
0\right\rangle _{\left(  0\right)  }=\frac{\partial\delta_{1}\varphi_{i}%
}{\partial c\left(  z^{\prime}\right)  }D^{\left(  0\right)  }\left(  \bar
{c}\left(  z\right)  ,c\left(  z^{\prime}\right)  \right)  \label{bpg2}%
\end{equation}
where the subscript and superscript $\left(  0\right)  $ refer to tree order
terms and $\frac{\partial\delta_{1}\varphi_{i}}{\partial c}$ is a constant or
constant operator. Next, let us assume that $\varphi_{j}$ and $\varphi_{k}$
are non-ghost fields. Then $\left(  \ref{bidt}\right)  $ yields%
\begin{align}
T\left\langle 0\left\vert \left(  \delta\bar{c}\left(  z\right)  \right)
\varphi_{j}\varphi_{k}\right\vert 0\right\rangle _{\left(  0\right)  }  &
=\ T\left\langle 0\left\vert \bar{c}\left(  z\right)  \left(  \delta
\varphi_{j}\right)  \varphi_{k}\right\vert 0\right\rangle _{\left(  0\right)
}\label{bvx3}\\
&  +\ T\left\langle 0\left\vert \bar{c}\left(  z\right)  \varphi_{j}\left(
\delta\varphi_{k}\right)  \right\vert 0\right\rangle _{\left(  0\right)
}\nonumber
\end{align}
According to the definition $\left(  \ref{udtp}\right)  $ for the Green
function with underlined arguments, the left side of $\left(  \ref{bvx3}%
\right)  $ may be expressed as%
\begin{align*}
T\left\langle 0\left\vert \left(  \delta\bar{c}\left(  z\right)  \right)
\varphi_{j}\varphi_{k}\right\vert 0\right\rangle _{\left(  0\right)  }  &
=D^{\left(  0\right)  }\left(  \delta\bar{c}\left(  z\right)  ,\varphi
_{i}\right)  G^{\left(  0\right)  }\left(  \underline{\varphi_{i}},\varphi
_{j},\varphi_{k}\right) \\
&  =D^{\left(  0\right)  }\left(  \bar{c}\left(  z\right)  ,c\left(
z^{\prime}\right)  \right)  \frac{\partial\delta_{1}\varphi_{i}}{\partial
c\left(  z^{\prime}\right)  }G^{\left(  0\right)  }\left(  \underline
{\varphi_{i}},\varphi_{j},\varphi_{k}\right)
\end{align*}
In the tree order, the anti-ghost $\bar{c}\left(  z\right)  $ field in
$T\left\langle 0\left\vert \bar{c}\left(  z\right)  \left(  \delta\varphi
_{j}\right)  \varphi_{k}\right\vert 0\right\rangle _{\left(  0\right)  }$,
which is the first term on the right side of $\left(  \ref{bvx3}\right)  $,
must be paired under Wick contraction with the ghost $c$ field in
$\delta\varphi_{j}$ or with the $c$ field from the interaction Lagrangian, and
we have%
\begin{align}
T\left\langle 0\left\vert \bar{c}\left(  z\right)  \left(  \delta\varphi
_{j}\right)  \varphi_{k}\right\vert 0\right\rangle _{\left(  0\right)  }  &
=D^{\left(  0\right)  }\left(  \bar{c}\left(  z\right)  ,c\left(  z^{\prime
}\right)  \right)  \times\label{gcp1}\\
&  \left[  D^{\left(  0\right)  }\left(  \frac{\partial\delta_{2}\varphi_{j}%
}{\partial c\left(  z^{\prime}\right)  },\varphi_{k}\right)  +G^{\left(
0\right)  }\left(  \underline{c\left(  z^{\prime}\right)  },\delta_{1}%
\varphi_{j},\varphi_{k}\right)  \right] \nonumber
\end{align}
with the Green function%
\begin{equation}
G^{\left(  0\right)  }\left(  \underline{c},c,\varphi_{k};k_{1},k_{2}%
,k_{3}\right)  =D^{\left(  0\right)  }\left(  c,\bar{c};k_{2}\right)
\Gamma^{\left(  0\right)  }\left(  c,\bar{c},\varphi_{i}\right)  D^{\left(
0\right)  }\left(  \varphi_{i},\varphi_{k};k_{3}\right)  \label{gvf}%
\end{equation}
where $\Gamma^{\left(  0\right)  }\left(  c,\bar{c},\varphi\right)  $
represents the vertex factor of $c-\bar{c}-\varphi$. Note that we have
discarded $D^{\left(  0\right)  }\left(  \frac{\partial\delta_{1}\varphi_{j}%
}{\partial c},\varphi_{k}\right)  $ owing to the vanishing vacuum expectation
$\left\langle \varphi_{k}\right\rangle =0$. For the 2nd term on the right side
of $\left(  \ref{bvx3}\right)  $, there is an expression similar to $\left(
\ref{gcp1}\right)  $. The identity $\left(  \ref{bvx3}\right)  $, after
factoring out the common ghost propagator $D^{\left(  0\right)  }\left(
\bar{c}\left(  z\right)  ,c\left(  z^{\prime}\right)  \right)  $ and then
replacing $z^{\prime}$ by $z$, becomes%
\begin{align}
\frac{\partial\delta_{1}\varphi_{i}}{\partial c\left(  z\right)  }G^{\left(
0\right)  }\left(  \underline{\varphi_{i}},\varphi_{j},\varphi_{k}\right)   &
=D^{\left(  0\right)  }\left(  \frac{\partial\delta_{2}\varphi_{j}}{\partial
c\left(  z\right)  },\varphi_{k}\right)  +D^{\left(  0\right)  }\left(
\varphi_{j}\left(  x\right)  ,\frac{\partial\delta_{2}\varphi_{k}}{\partial
c\left(  z\right)  }\right) \label{bvx3a}\\
&  +G^{\left(  0\right)  }\left(  \underline{c\left(  z\right)  },\delta
_{1}\varphi_{j},\varphi_{k}\right)  +G^{\left(  0\right)  }\left(
\underline{c\left(  z\right)  },\varphi_{j},\delta_{1}\varphi_{k}\right)
\nonumber
\end{align}
The definition $\left(  \ref{defeta}\right)  $ for the composite vertex $%
\begin{picture}(12,12) (10,-8)
\SetWidth{0.5}
\SetColor{Black}
\COval(16,-4)(5.657,5.657)(45.0){Black}{White}\Line
(13.172,-6.828)(18.828,-1.172)\Line(13.172,-1.172)(18.828,-6.828)
\end{picture}%
$ on a fermion line may be extended to include other types of vertices. The
extended composite vertex is defined as%
\begin{align}%
\begin{picture}(12,12) (10,-8)
\SetWidth{0.5}
\SetColor{Black}
\COval(16,-4)(5.657,5.657)(45.0){Black}{White}\Line
(13.172,-6.828)(18.828,-1.172)\Line(13.172,-1.172)(18.828,-6.828)
\end{picture}%
&  =\frac{\partial\delta_{1}\varphi_{i}}{\partial c}\Gamma^{\left(  0\right)
}\left(  \varphi_{i},\varphi,\varphi^{\prime}\right) \label{extcr}\\
&  =-ik_{\mu}\Gamma^{\left(  0\right)  }\left(  A^{\mu},\varphi,\varphi
^{\prime}\right)  -M\Gamma^{\left(  0\right)  }\left(  \phi_{2},\varphi
,\varphi^{\prime}\right) \nonumber
\end{align}
where the tree order amplitude $\Gamma^{\left(  0\right)  }\left(  \varphi
_{i},\varphi,\varphi^{\prime}\right)  $ stands for the vertex factor of
$\varphi_{i}-\varphi-\varphi^{\prime}$ and $k$ is the incoming momentum of the
vector field $A^{\mu}$ or scalar field $\phi_{2}$. Note that this definition
is the same as the restricted one of $\left(  \ref{defeta}\right)  $ when
$\varphi$ and $\varphi^{\prime}$ are the fermion fields $\psi$ and $\bar{\psi
}$. The amplitude $\frac{\partial\delta_{1}\varphi_{i}}{\partial c}G^{\left(
0\right)  }\left(  \underline{\varphi_{i}},\varphi_{j},\varphi_{k}\right)  $
can then be diagrammatically expressed as a composite vertex $%
\begin{picture}(12,12) (10,-8)
\SetWidth{0.5}
\SetColor{Black}
\COval(16,-4)(5.657,5.657)(45.0){Black}{White}\Line
(13.172,-6.828)(18.828,-1.172)\Line(13.172,-1.172)(18.828,-6.828)
\end{picture}%
$ connected with two propagator lines to fields $\varphi_{j}$ and $\varphi
_{k}$:
\begin{equation}%
\raisebox{-16pt}{
\begin{picture}(56,34) (-7,-10)
\SetWidth{0.5}
\SetColor{Black}
\Line(32,18)(16,2)
\Line(32,18)(48,2)
\COval(32,18)(5.657,5.657)(45.0){Black}{White}\Line
(29.172,15.172)(34.828,20.828)\Line(29.172,20.828)(34.828,15.172)
\Text(16,-4)[]{\normalsize{\Black{$\varphi_j$}}}
\Text(48,-4)[]{\normalsize{\Black{$\varphi_k$}}}
\end{picture}
}%
=\frac{\partial\delta_{1}\varphi_{i}}{\partial c}G^{\left(  0\right)  }\left(
\underline{\varphi_{i}},\varphi_{j},\varphi_{k}\right)  \label{cpv0}%
\end{equation}

Let us use a solid black box
\begin{picture}(5,5) (0,-1)
\SetWidth{0.5}
\SetColor{Black}
\CBox(5.83,-0.83)(0.17,4.83){Black}{Black}
\end{picture}
\ to graphically represent the $c-\bar{c}-\varphi$ vertex. Then the Green
function $\left(  \ref{gvf}\right)  $ can be diagrammatically expressed as%
\begin{equation}%
\begin{picture}(60,20) (5,-6)
\SetWidth{0.5}
\SetColor{Black}
\CBox(44.828,5.172)(39.172,10.828){Black}{Black}
\Line
[dash,dashsize=2,arrow,arrowpos=0.5,arrowlength=2.5,arrowwidth=1,arrowinset=0.2](40,8)(20,-2)
\Line[dash,dashsize=0.5](60,-2)(40,8)
\Text(20,-10)[]{\normalsize{\Black{$c$}}}
\Text(60,-10)[]{\normalsize{\Black{$\varphi_i$}}}
\end{picture}%
=G^{\left(  0\right)  }\left(  \underline{c},c,\varphi_{i}\right)
\label{g-02}%
\end{equation}
where the dotted arrowed line corresponds to the ghost propagator $D^{\left(
0\right)  }\left(  c,\bar{c}\right)  $. Let us also define%
\begin{equation}%
\begin{picture}(60,22) (5,-6)
\SetWidth{0.5}
\SetColor{Black}
\CBox(44.828,7.172)(39.172,12.828){Black}{Black}
\Line
[dash,dashsize=2,arrow,arrowpos=0.5,arrowlength=2.5,arrowwidth=1,arrowinset=0.2](40,10)(20,0)
\Line[dash,dashsize=0.5](60,0)(40,10)
\Text(20,-10)[]{\normalsize{\Black{$\delta_1 \varphi_j$}}}
\Text(60,-10)[]{\normalsize{\Black{$\varphi_i$}}}
\COval(20,0)(4.243,4.243)(45.0){Black}{White}\Line
(17.879,-2.121)(22.121,2.121)\Line(17.879,2.121)(22.121,-2.121)
\end{picture}%
=G^{\left(  0\right)  }\left(  \underline{c},\delta_{1}\varphi_{j},\varphi
_{i}\right)  =\frac{\partial\delta_{1}\varphi_{j}}{\partial c}G^{\left(
0\right)  }\left(  \underline{c},c,\varphi_{i}\right)  \label{g-02-2}%
\end{equation}

Since $\delta_{2}\varphi_{j}$ is a product of one ghost $c$ field and another
non-ghost field, taking the partial derivative with respect to $c$ as in
$\frac{\partial\delta_{2}\varphi_{j}}{\partial c}$ is equivalent to factoring
out the $c$ field to retain the non-ghost factor. $D^{\left(  0\right)
}\left(  \frac{\partial\delta_{2}\varphi_{j}}{\partial c},\varphi_{k}\right)
$ is thus proportional to the free propagator that propagates the field
$\varphi_{k}$ to the non-ghost field in $\delta_{2}\varphi_{j}$. In
particular, if $\varphi_{j}=\phi_{2}$ and $\varphi_{k}=H$, then $\delta
_{2}\phi_{2}=-gcH$ and%
\[
D^{\left(  0\right)  }\left(  \frac{\partial\delta_{2}\phi_{2}}{\partial
c},H\right)  =-gD^{\left(  0\right)  }\left(  H,H\right)  .
\]
We now graphically represent $D^{\left(  0\right)  }\left(  \frac
{\partial\delta_{2}\varphi_{j}}{\partial c},\varphi_{k}\right)  $ by%
\begin{equation}%
\begin{picture}(64,34) (1,-10)
\SetWidth{0.5}
\SetColor{Black}
\Line
[arrow,arrowpos=1,arrowlength=2.5,arrowwidth=1,arrowinset=0.2](30,22)(14,6)
\Line(32,18)(16,2)
\Line(32,18)(48,2)
\COval(32,18)(5.657,5.657)(45.0){Black}{White}\Line
(29.172,15.172)(34.828,20.828)\Line(29.172,20.828)(34.828,15.172)
\Text(16,-4)[]{\normalsize{\Black{$\delta_2 \varphi_j$}}}
\Text(48,-4)[]{\normalsize{\Black{$\varphi_k$}}}
\end{picture}
\label{wtic1-0a}%
\end{equation}
where the single line stands for the free propagator from $\varphi_{k}$ to the
non-ghost field in $\delta_{2}\varphi_{j}$ \ and the arrowed double line
emitting from the composite vertex $%
\begin{picture}(12,12) (10,-8)
\SetWidth{0.5}
\SetColor{Black}
\COval(16,-4)(5.657,5.657)(45.0){Black}{White}\Line
(13.172,-6.828)(18.828,-1.172)\Line(13.172,-1.172)(18.828,-6.828)
\end{picture}%
$ is interpreted as that the original propagator connecting to field
$\varphi_{j}$ as in $\left(  \ref{cpv0}\right)  $ is annihilated and the
composite vertex with the arrowed double line is to be replaced by the
constant coefficient of the non-ghost field in $\frac{\partial\delta
_{2}\varphi_{k}}{\partial c}$. With the graphical elements defined in $\left(
\text{\ref{extcr}}\right)  $-$\left(  \text{\ref{wtic1-0a}}\right)  $, the
identity $\left(  \ref{bvx3a}\right)  $ can be diagrammatically expressed as%
\begin{equation}%
\begin{picture}(307,36) (-5,-10)
\SetWidth{0.5}
\SetColor{Black}
\Line
[arrow,arrowpos=1,arrowlength=2.5,arrowwidth=1,arrowinset=0.2](152,19)(170,10)
\Line
[arrow,arrowpos=1,arrowlength=2.5,arrowwidth=1,arrowinset=0.2](88,19)(70,10)
\Text(120,16)[]{\normalsize{\Black{$+$}}}
\Text(60,16)[]{\normalsize{\Black{$=$}}}
\Text(10,-4)[]{\normalsize{\Black{$\varphi_j$}}}
\Text(50,-4)[]{\normalsize{\Black{$\varphi_k$}}}
\Text(75,-4)[]{\normalsize{\Black{$\delta_2 \varphi_j$}}}
\Text(110,-4)[]{\normalsize{\Black{$\varphi_k$}}}
\Text(130,-4)[]{\normalsize{\Black{$\varphi_j$}}}
\Text(165,-4)[]{\normalsize{\Black{$\delta_2 \varphi_k$}}}
\CBox(272.828,13.172)(267.172,18.828){Black}{Black}
\Text(240,16)[]{\normalsize{\Black{$+$}}}
\Text(250,-4)[]{\normalsize{\Black{$\varphi_j$}}}
\Text(285,-4)[]{\normalsize{\Black{$\delta_1 \varphi_k$}}}
\Line(270,16)(250,6)
\Line
[dash,dashsize=2,arrow,arrowpos=0.5,arrowlength=2.5,arrowwidth=1,arrowinset=0.2](270,16)(290,6)
\COval(290,6)(4.243,4.243)(45.0){Black}{White}\Line
(287.879,3.879)(292.121,8.121)\Line(287.879,8.121)(292.121,3.879)
\CBox(212.828,13.172)(207.172,18.828){Black}{Black}
\Text(180,16)[]{\normalsize{\Black{$+$}}}
\Text(195,-4)[]{\normalsize{\Black{$\delta_1 \varphi_j$}}}
\Text(230,-4)[]{\normalsize{\Black{$\varphi_k$}}}
\Line
[dash,dashsize=2,arrow,arrowpos=0.5,arrowlength=2.5,arrowwidth=1,arrowinset=0.2](210,16)(190,6)
\Line(210,16)(230,6)
\COval(190,6)(4.243,4.243)(45.0){Black}{White}\Line
(187.879,3.879)(192.121,8.121)\Line(187.879,8.121)(192.121,3.879)
\Line(30,16)(10,6)
\Line(90,16)(70,6)
\Line(150,16)(130,6)
\Line(90,16)(110,6)
\Line(150,16)(170,6)
\Line(30,16)(50,6)
\COval(30,16)(4.243,4.243)(45.0){Black}{White}\Line
(27.879,13.879)(32.121,18.121)\Line(27.879,18.121)(32.121,13.879)
\COval(90,16)(4.243,4.243)(45.0){Black}{White}\Line
(87.879,13.879)(92.121,18.121)\Line(87.879,18.121)(92.121,13.879)
\COval(150,16)(4.243,4.243)(45.0){Black}{White}\Line
(147.879,13.879)(152.121,18.121)\Line(147.879,18.121)(152.121,13.879)
\end{picture}
\label{b-17}%
\end{equation}
Likewise, by expanding%
\[
T\left\langle 0\left\vert \delta\left(  \bar{c}\left(  z\right)  \varphi
_{i}\varphi_{j}\varphi_{k}\right)  \right\vert 0\right\rangle _{\left(
0\right)  }=0
\]
and utilizing $\left(  \ref{b-17}\right)  $, we get the identity%
\begin{equation}%
\begin{picture}(262,66) (-5,0)
\SetWidth{0.5}
\SetColor{Black}
\Line
[arrow,arrowpos=1,arrowlength=2.5,arrowwidth=1,arrowinset=0.2](78,30)(88,40)
\Line
[arrow,arrowpos=1,arrowlength=2.5,arrowwidth=1,arrowinset=0.2](26,46)(26,36)
\Line(30,56)(30,36)
\Line(30,36)(10,16)
\Line(30,36)(50,16)
\Text(40,56)[]{\normalsize{\Black{$\varphi_k$}}}
\Text(50,6)[]{\normalsize{\Black{$\varphi_j$}}}
\Text(10,6)[]{\normalsize{\Black{$\varphi_i$}}}
\Line(90,56)(90,36)
\Line(90,36)(70,16)
\Line(90,36)(110,16)
\Text(100,56)[]{\normalsize{\Black{$\varphi_k$}}}
\Text(110,6)[]{\normalsize{\Black{$\varphi_j$}}}
\Text(70,6)[]{\normalsize{\Black{$\varphi_i$}}}
\Line(150,56)(150,36)
\Line(150,36)(130,16)
\Line(150,36)(170,16)
\Text(160,56)[]{\normalsize{\Black{$\varphi_k$}}}
\Text(170,6)[]{\normalsize{\Black{$\varphi_j$}}}
\Text(130,6)[]{\normalsize{\Black{$\varphi_i$}}}
\Line(210,56)(210,36)
\Line(210,36)(190,16)
\Line(210,36)(230,16)
\Text(220,56)[]{\normalsize{\Black{$\varphi_k$}}}
\Text(230,6)[]{\normalsize{\Black{$\varphi_j$}}}
\Text(190,6)[]{\normalsize{\Black{$\varphi_i$}}}
\COval(210,36)(4.243,4.243)(45.0){Black}{White}\Line
(207.879,33.879)(212.121,38.121)\Line(207.879,38.121)(212.121,33.879)
\COval(30,46)(4.243,4.243)(45.0){Black}{White}\Line
(27.879,43.879)(32.121,48.121)\Line(27.879,48.121)(32.121,43.879)
\COval(80,26)(4.243,4.243)(45.0){Black}{White}\Line
(77.879,23.879)(82.121,28.121)\Line(77.879,28.121)(82.121,23.879)
\COval(160,26)(4.243,4.243)(45.0){Black}{White}\Line
(157.879,23.879)(162.121,28.121)\Line(157.879,28.121)(162.121,23.879)
\Text(240,36)[]{\normalsize{\Black{$= \ 0$}}}
\Text(60,36)[]{\normalsize{\Black{$+$}}}
\Text(120,36)[]{\normalsize{\Black{$+$}}}
\Text(180,36)[]{\normalsize{\Black{$+$}}}
\Line
[arrow,arrowpos=1,arrowlength=2.5,arrowwidth=1,arrowinset=0.2](162,30)(152,40)
\end{picture}
\label{wtic1-3}%
\end{equation}
The above two graphic identities $\left(  \ref{b-17}\right)  $ and $\left(
\ref{wtic1-3}\right)  $ together with the condition%
\[
T\left\langle 0\left\vert \delta\left(  \bar{c}\left(  z\right)  \varphi
_{i}\varphi_{j}\varphi_{k}\varphi_{l}\right)  \right\vert 0\right\rangle
_{\left(  0\right)  }=0
\]
can be combined to yield the identity%
\begin{equation}%
\begin{picture}(262,76) (15,0)
\SetWidth{0.5}
\SetColor{Black}
\Line
[arrow,arrowpos=1,arrowlength=2.5,arrowwidth=1,arrowinset=0.2](38,30)(45,37)
\Line
[arrow,arrowpos=1,arrowlength=2.5,arrowwidth=1,arrowinset=0.2](122,30)(115,37)
\Line
[arrow,arrowpos=1,arrowlength=2.5,arrowwidth=1,arrowinset=0.2](178,50)(169,41)
\Line
[arrow,arrowpos=1,arrowlength=2.5,arrowwidth=1,arrowinset=0.2](222,50)(231,41)
\Text(200,36)[]{\normalsize{\Black{$+$}}}
\Line(250,56)(210,16)
\Line(210,56)(250,16)
\Text(250,66)[]{\normalsize{\Black{$\varphi_k$}}}
\Text(250,6)[]{\normalsize{\Black{$\varphi_j$}}}
\Text(210,6)[]{\normalsize{\Black{$\varphi_i$}}}
\Text(210,66)[]{\normalsize{\Black{$\varphi_l$}}}
\Text(260,36)[]{\normalsize{\Black{$= \ 0$}}}
\Text(80,36)[]{\normalsize{\Black{$+$}}}
\Text(140,36)[]{\normalsize{\Black{$+$}}}
\Text(70,66)[]{\normalsize{\Black{$\varphi_k$}}}
\Text(70,6)[]{\normalsize{\Black{$\varphi_j$}}}
\Text(30,6)[]{\normalsize{\Black{$\varphi_i$}}}
\Text(130,66)[]{\normalsize{\Black{$\varphi_k$}}}
\Text(130,6)[]{\normalsize{\Black{$\varphi_j$}}}
\Text(90,6)[]{\normalsize{\Black{$\varphi_i$}}}
\Text(190,66)[]{\normalsize{\Black{$\varphi_k$}}}
\Text(190,6)[]{\normalsize{\Black{$\varphi_j$}}}
\Text(150,6)[]{\normalsize{\Black{$\varphi_i$}}}
\Line(30,56)(70,16)
\Line(90,56)(130,16)
\Line(150,56)(190,16)
\Line(70,56)(30,16)
\Line(130,56)(90,16)
\Line(190,56)(150,16)
\Text(30,66)[]{\normalsize{\Black{$\varphi_l$}}}
\Text(90,66)[]{\normalsize{\Black{$\varphi_l$}}}
\Text(150,66)[]{\normalsize{\Black{$\varphi_l$}}}
\COval(220,46)(4.243,4.243)(45.0){Black}{White}\Line
(217.879,43.879)(222.121,48.121)\Line(217.879,48.121)(222.121,43.879)
\COval(180,46)(4.243,4.243)(45.0){Black}{White}\Line
(177.879,43.879)(182.121,48.121)\Line(177.879,48.121)(182.121,43.879)
\COval(120,26)(4.243,4.243)(45.0){Black}{White}\Line
(117.879,23.879)(122.121,28.121)\Line(117.879,28.121)(122.121,23.879)
\COval(40,26)(4.243,4.243)(45.0){Black}{White}\Line
(37.879,23.879)(42.121,28.121)\Line(37.879,28.121)(42.121,23.879)
\end{picture}
\label{wtic1-4}%
\end{equation}

We will need graphical notations to express two amputated external fields in a
four-point function. In the following figure%
\begin{equation}%
\begin{picture}(72,39) (-5,-13)
\SetWidth{0.5}
\SetColor{Black}
\Line[dash,dashsize=0.5](30,9)(10,-1)
\Line[dash,dashsize=0.5](30,9)(50,-1)
\Line(33,19)(27,13)\Line(33,13)(27,19)
\Text(10,-7)[]{\normalsize{\Black{$\varphi_i$}}}
\Text(50,-7)[]{\normalsize{\Black{$\varphi_j$}}}
\Text(40,16)[]{\normalsize{\Black{$\mu$}}}
\Line(33,12)(27,6)\Line(33,6)(27,12)
\Text(40,9)[]{\normalsize{\Black{$\nu$}}}
\end{picture}
\label{wtic1-6}%
\end{equation}
the amputated $A^{\mu}$ and $A^{\nu}$\ fields are represented by two crosses
that are stacked together. Similarly,
\begin{equation}%
\begin{picture}(72,39) (-5,-13)
\SetWidth{0.5}
\SetColor{Black}
\Line[dash,dashsize=0.5](30,9)(10,-1)
\Line[dash,dashsize=0.5](30,9)(50,-1)
\COval(30,9)(4.243,4.243)(45.0){Black}{White}\Line
(27.879,6.879)(32.121,11.121)\Line(27.879,11.121)(32.121,6.879)
\Line(33,19)(27,13)\Line(33,13)(27,19)
\Text(10,-7)[]{\normalsize{\Black{$\varphi_i$}}}
\Text(50,-7)[]{\normalsize{\Black{$\varphi_j$}}}
\Text(40,16)[]{\normalsize{\Black{$\mu$}}}
\end{picture}
\label{wtic1-5}%
\end{equation}
represents a four-point function with an amputated external $A^{\mu}$ and a
composite vertex $%
\begin{picture}(12,12) (10,-8)
\SetWidth{0.5}
\SetColor{Black}
\COval(16,-4)(5.657,5.657)(45.0){Black}{White}\Line
(13.172,-6.828)(18.828,-1.172)\Line(13.172,-1.172)(18.828,-6.828)
\end{picture}%
$.

We are now equipped with the graphical notations and identities needed to
construct component diagrams for Ward identities without the restriction on
the type of internal field lines.

\subsection{Two-Loop Triangular Ward Identity}

If all the vertices for external fields are detached, a 3-point 2-loop 1PI
diagram in the presence of one fermion-loop sub-diagram becomes a 2-loop
super-generator diagram%
\begin{equation}%
\begin{picture}(34,32) (23,-9)
\SetWidth{0.5}
\SetColor{Black}
\Line[dash,dashsize=0.5](40,22)(40,-8)
\Arc
[arrow,arrowpos=0.5,arrowlength=2.5,arrowwidth=1,arrowinset=0.2](40,7)(15,90,270)
\Arc
[arrow,arrowpos=0.5,arrowlength=2.5,arrowwidth=1,arrowinset=0.2](40,7)(15,-90,90)
\end{picture}
\label{a-03-2}%
\end{equation}
which is composed of a fermion loop and a non-fermion internal line. Seven
topologically different generator diagrams will result from all possible
attachments of the vertices for $A^{\mu}$ and $A^{\nu}$ consistent with
Feynman rules to this super-generator:%
\begin{equation}%
\begin{picture}(261,83) (6,-28)
\SetWidth{0.5}
\SetColor{Black}
\Arc
[arrow,arrowpos=0.5,arrowlength=2.5,arrowwidth=1,arrowinset=0.2](110,38)(15,-0,180)
\Arc
[arrow,arrowpos=0.5,arrowlength=2.5,arrowwidth=1,arrowinset=0.2](110,38)(15,-180,0)
\Arc
[arrow,arrowpos=0.5,arrowlength=2.5,arrowwidth=1,arrowinset=0.2](40,38)(15,-0,180)
\Arc
[arrow,arrowpos=0.5,arrowlength=2.5,arrowwidth=1,arrowinset=0.2](40,38)(15,-180,0)
\Arc[dash,dashsize=0.5,clock](40,54.333)(11.333,-28.072,-151.928)
\Arc[dash,dashsize=0.5,clock](110,54.333)(11.333,-28.072,-151.928)
\Line[dash,dashsize=0.5](40,3)(40,-27)
\Arc
[arrow,arrowpos=0.5,arrowlength=2.5,arrowwidth=1,arrowinset=0.2](40,-12)(15,-90,90)
\Arc
[arrow,arrowpos=0.5,arrowlength=2.5,arrowwidth=1,arrowinset=0.2](40,-12)(15,90,270)
\Arc
[arrow,arrowpos=0.25,arrowlength=2.5,arrowwidth=1,arrowinset=0.2](180,38)(15,-0,180)
\Arc
[arrow,arrowpos=0.25,arrowlength=2.5,arrowwidth=1,arrowinset=0.2](180,38)(15,-180,0)
\Line[dash,dashsize=0.5](180,53)(180,23)
\Arc
[arrow,arrowpos=0.25,arrowlength=2.5,arrowwidth=1,arrowinset=0.2](110,-12)(15,90,270)
\Line
[arrow,arrowpos=0.5,arrowlength=2.5,arrowwidth=1,arrowinset=0.2](110,-27)(110,3)
\Arc[dash,dashsize=0.5](110,-12)(15,-90,90)
\Arc[dash,dashsize=0.5](180,-12)(15,-90,90)
\Line
[arrow,arrowpos=0.5,arrowlength=2.5,arrowwidth=1,arrowinset=0.2](180,-27)(180,3)
\Arc
[arrow,arrowpos=0.25,arrowlength=2.5,arrowwidth=1,arrowinset=0.2](180,-12)(15,90,270)
\Text(250,43)[]{\normalsize{\Black{$\mu$}}}
\Line(237,40)(243,46)\Line(237,46)(243,40)
\Line[dash,dashsize=0.5](240,53)(240,23)
\Arc
[arrow,arrowpos=0.5,arrowlength=2.5,arrowwidth=1,arrowinset=0.2](240,38)(15,-90,90)
\Arc
[arrow,arrowpos=0.5,arrowlength=2.5,arrowwidth=1,arrowinset=0.2](240,38)(15,90,270)
\Text(21,30)[]{\normalsize{\Black{$\mu$}}}
\Line(22,35)(28,41)\Line(22,41)(28,35)
\Text(59,30)[]{\normalsize{\Black{$\nu$}}}
\Line(52,35)(58,41)\Line(52,41)(58,35)
\Text(91,30)[]{\normalsize{\Black{$\nu$}}}
\Line(92,35)(98,41)\Line(92,41)(98,35)
\Text(129,30)[]{\normalsize{\Black{$\mu$}}}
\Line(122,35)(128,41)\Line(122,41)(128,35)
\Text(45,-5)[]{\normalsize{\Black{$\nu$}}}
\Text(35,-5)[]{\normalsize{\Black{$\mu$}}}
\Line(33,-15)(39,-9)\Line(33,-9)(39,-15)
\Line(41,-15)(47,-9)\Line(41,-9)(47,-15)
\Text(250,33)[]{\normalsize{\Black{$\nu$}}}
\Line(237,30)(243,36)\Line(237,36)(243,30)
\Text(199,30)[]{\normalsize{\Black{$\nu$}}}
\Line(192,35)(198,41)\Line(192,41)(198,35)
\Text(161,30)[]{\normalsize{\Black{$\mu$}}}
\Line(162,35)(168,41)\Line(162,41)(168,35)
\Text(199,-20)[]{\normalsize{\Black{$\nu$}}}
\Line(192,-15)(198,-9)\Line(192,-9)(198,-15)
\Text(161,-20)[]{\normalsize{\Black{$\mu$}}}
\Line(162,-15)(168,-9)\Line(162,-9)(168,-15)
\Text(129,-20)[]{\normalsize{\Black{$\mu$}}}
\Line(122,-15)(128,-9)\Line(122,-9)(128,-15)
\Text(91,-20)[]{\normalsize{\Black{$\nu$}}}
\Line(92,-15)(98,-9)\Line(92,-9)(98,-15)
\end{picture}
\label{a-03-1}%
\end{equation}
For each diagram in the above, either of the two legitimate cut points at the
two vertices connecting to the non-fermion internal line is available to yield
a cut generator for a regularized Ward identity. For example, if the cutting
is made at the endpoint of the internal fermion line connecting to the lowest
vertex on the last diagram in $\left(  \ref{a-03-1}\right)  $, we obtain the
generator%
\begin{equation}%
\begin{picture}(60,39) (5,-7)
\SetWidth{0.5}
\SetColor{Black}
\Text(48,22)[]{\normalsize{\Black{$\nu$}}}
\Text(20,-1)[]{\normalsize{\Black{$\mu$}}}
\SetWidth{0.5}
\Line(50,9)(10,9)
\Line(17,6)(23,12)\Line(17,12)(23,6)
\Line(37,16)(43,22)\Line(37,22)(43,16)
\Arc[dash,dashsize=0.5](40,9)(10,-0,180)
\end{picture}
\label{b-20}%
\end{equation}
The vertex for $A^{\mu}$ is attached to the fermion line and the vertex for
$A^{\nu}$ is attached to the arc above the fermion line. We may attach $%
\begin{picture}(12,12) (10,-8)
\SetWidth{0.5}
\SetColor{Black}
\COval(16,-4)(5.657,5.657)(45.0){Black}{White}\Line
(13.172,-6.828)(18.828,-1.172)\Line(13.172,-1.172)(18.828,-6.828)
\end{picture}%
$ to the above cut generator $\left(  \ref{b-20}\right)  $ in all possible
manners to obtain the following collection of component diagrams:%
\begin{equation}%
\begin{picture}(200,79) (5,-7)
\SetWidth{0.5}
\SetColor{Black}
\Arc[dash,dashsize=0.5](45,4.273)(15.727,17.492,162.508)
\Arc[dash,dashsize=0.5](115,4.273)(15.727,17.492,162.508)
\Text(110,26)[]{\normalsize{\Black{$\nu$}}}
\Text(90,-1)[]{\normalsize{\Black{$\mu$}}}
\Line(107,16)(113,22)\Line(107,22)(113,16)
\Text(58,22)[]{\normalsize{\Black{$\nu$}}}
\Text(20,-1)[]{\normalsize{\Black{$\mu$}}}
\Line(60,9)(10,9)
\Line(47,16)(53,22)\Line(47,22)(53,16)
\Text(160,-1)[]{\normalsize{\Black{$\mu$}}}
\Line(190,9)(150,9)
\Line(157,6)(163,12)\Line(157,12)(163,6)
\Arc[dash,dashsize=0.5](180,9)(10,-0,180)
\COval(180,19)(4.243,4.243)(45.0){Black}{White}\Line
(177.879,16.879)(182.121,21.121)\Line(177.879,21.121)(182.121,16.879)
\Line(130,9)(80,9)
\COval(120,19)(4.243,4.243)(45.0){Black}{White}\Line
(117.879,16.879)(122.121,21.121)\Line(117.879,21.121)(122.121,16.879)
\Line(87,6)(93,12)\Line(87,12)(93,6)
\Line(17,6)(23,12)\Line(17,12)(23,6)
\Text(128,62)[]{\normalsize{\Black{$\nu$}}}
\Text(90,39)[]{\normalsize{\Black{$\mu$}}}
\Line(117,56)(123,62)\Line(117,62)(123,56)
\Arc[dash,dashsize=0.5](120,49)(10,-0,180)
\Text(58,62)[]{\normalsize{\Black{$\nu$}}}
\Text(30,39)[]{\normalsize{\Black{$\mu$}}}
\Line(60,49)(10,49)
\Line(47,56)(53,62)\Line(47,62)(53,56)
\Arc[dash,dashsize=0.5](50,49)(10,-0,180)
\Text(188,62)[]{\normalsize{\Black{$\nu$}}}
\Text(160,39)[]{\normalsize{\Black{$\mu$}}}
\Line(190,49)(150,49)
\Line(157,46)(163,52)\Line(157,52)(163,46)
\Line(177,56)(183,62)\Line(177,62)(183,56)
\Arc[dash,dashsize=0.5](180,49)(10,-0,180)
\COval(180,49)(4.243,4.243)(45.0){Black}{White}\Line
(177.879,46.879)(182.121,51.121)\Line(177.879,51.121)(182.121,46.879)
\Line(130,49)(80,49)
\COval(100,49)(4.243,4.243)(45.0){Black}{White}\Line
(97.879,46.879)(102.121,51.121)\Line(97.879,51.121)(102.121,46.879)
\Line(87,46)(93,52)\Line(87,52)(93,46)
\COval(20,49)(4.243,4.243)(45.0){Black}{White}\Line
(17.879,46.879)(22.121,51.121)\Line(17.879,51.121)(22.121,46.879)
\Line(27,46)(33,52)\Line(27,52)(33,46)
\COval(40,19)(4.243,4.243)(45.0){Black}{White}\Line
(37.879,16.879)(42.121,21.121)\Line(37.879,21.121)(42.121,16.879)
\Text(188,29)[]{\normalsize{\Black{$\nu$}}}
\Line(177,23)(183,29)\Line(177,29)(183,23)
\end{picture}
\label{b-21}%
\end{equation}
A component diagram is constructed when we insert $%
\begin{picture}(12,12) (10,-8)
\SetWidth{0.5}
\SetColor{Black}
\COval(16,-4)(5.657,5.657)(45.0){Black}{White}\Line
(13.172,-6.828)(18.828,-1.172)\Line(13.172,-1.172)(18.828,-6.828)
\end{picture}%
$ in consistency with Feynman rules into one of the internal lines or vertices
in the generator. The momentum entering the open fermion line from the right
side is assumed to be equal to the momentum leaving the fermion line at the
left end. Since the original closed fermion loop is restored by fusing the
open fermion line, the amplitude of the cut diagram is calculated by taking
the trace and carrying out the fermion-loop momentum integration.

There are many cancellations for the sum of component diagrams constructed
from a cut generator. Making use of $\left(  \ref{b-17}\right)  $ and $\left(
\ref{wtic1-3}\right)  $, the sum of the six diagrams in $\left(
\ref{b-21}\right)  $ becomes%
\begin{equation}%
\begin{picture}(262,116) (9,-10)
\SetWidth{0.5}
\SetColor{Black}
\Line
[arrow,arrowpos=1,arrowlength=2.5,arrowwidth=1,arrowinset=0.2](130,69)(122,69)
\Line
[arrow,arrowpos=1,arrowlength=2.5,arrowwidth=1,arrowinset=0.2](20,69)(12,69)
\Text(70,66)[]{\normalsize{\Black{$-$}}}
\Text(90,56)[]{\normalsize{\Black{$\mu$}}}
\Arc[dash,dashsize=0.5](110,66)(10,-0,180)
\Text(58,79)[]{\normalsize{\Black{$\nu$}}}
\Text(30,56)[]{\normalsize{\Black{$\mu$}}}
\Line(60,66)(10,66)
\Line(47,73)(53,79)\Line(47,79)(53,73)
\Arc[dash,dashsize=0.5](50,66)(10,-0,180)
\Line(130,66)(80,66)
\COval(130,66)(4.243,4.243)(45.0){Black}{White}\Line
(127.879,63.879)(132.121,68.121)\Line(127.879,68.121)(132.121,63.879)
\Line(87,63)(93,69)\Line(87,69)(93,63)
\COval(20,66)(4.243,4.243)(45.0){Black}{White}\Line
(17.879,63.879)(22.121,68.121)\Line(17.879,68.121)(22.121,63.879)
\Line(27,63)(33,69)\Line(27,69)(33,63)
\Text(118,79)[]{\normalsize{\Black{$\nu$}}}
\Line(107,73)(113,79)\Line(107,79)(113,73)
\Text(220,56)[]{\normalsize{\Black{$\mu$}}}
\Text(120,6)[]{\normalsize{\Black{$+$}}}
\Text(108,26)[]{\normalsize{\Black{$\nu$}}}
\Line(97,20)(103,26)\Line(97,26)(103,20)
\CBox(82.828,13.172)(77.172,18.828){Black}{Black}
\Line[dash,dashsize=0.5](80,16)(70,6)
\Line[dash,dashsize=0.5](100,16)(110,6)
\Line
[dash,dashsize=2,arrow,arrowpos=0.5,arrowlength=2.5,arrowwidth=1,arrowinset=0.2](80,16)(100,16)
\Line(110,6)(50,6)
\Line(57,3)(63,9)\Line(57,9)(63,3)
\Text(60,-4)[]{\normalsize{\Black{$\mu$}}}
\COval(100,16)(4.243,4.243)(45.0){Black}{White}\Line
(97.879,13.879)(102.121,18.121)\Line(97.879,18.121)(102.121,13.879)
\Text(40,6)[]{\normalsize{\Black{$+$}}}
\Line[dash,dashsize=0.5](180,6)(170,16)
\Line[dash,dashsize=0.5](170,16)(160,16)
\Text(190,6)[]{\normalsize{\Black{$+$}}}
\Text(210,-4)[]{\normalsize{\Black{$\mu$}}}
\Text(178,19)[]{\normalsize{\Black{$\nu$}}}
\Text(140,-4)[]{\normalsize{\Black{$\mu$}}}
\Line(180,6)(130,6)
\Line(167,13)(173,19)\Line(167,19)(173,13)
\Line(250,6)(200,6)
\COval(250,6)(4.243,4.243)(45.0){Black}{White}\Line
(247.879,3.879)(252.121,8.121)\Line(247.879,8.121)(252.121,3.879)
\Line(207,3)(213,9)\Line(207,9)(213,3)
\Line(137,3)(143,9)\Line(137,9)(143,3)
\Line
[dash,dashsize=2,arrow,arrowpos=0.5,arrowlength=2.5,arrowwidth=1,arrowinset=0.2](160,16)(150,6)
\CBox(162.828,13.172)(157.172,18.828){Black}{Black}
\COval(150,6)(4.243,4.243)(45.0){Black}{White}\Line
(147.879,3.879)(152.121,8.121)\Line(147.879,8.121)(152.121,3.879)
\Line[dash,dashsize=0.5](230,16)(220,6)
\Line[dash,dashsize=0.5](240,16)(230,16)
\Line
[dash,dashsize=2,arrow,arrowpos=0.5,arrowlength=2.5,arrowwidth=1,arrowinset=0.2](240,16)(250,6)
\Text(222,19)[]{\normalsize{\Black{$\nu$}}}
\Line(227,13)(233,19)\Line(227,19)(233,13)
\CBox(242.828,13.172)(237.172,18.828){Black}{Black}
\CBox(262.828,73.172)(257.172,78.828){Black}{Black}
\Line
[dash,dashsize=10,arrow,arrowpos=0.5,arrowlength=2.5,arrowwidth=1,arrowinset=0.2](260,76)(240,76)
\Line[dash,dashsize=0.5](240,76)(230,66)
\Line[dash,dashsize=0.5](260,76)(270,66)
\Line(270,66)(210,66)
\Line(217,63)(223,69)\Line(217,69)(223,63)
\COval(240,76)(4.243,4.243)(45.0){Black}{White}\Line
(237.879,73.879)(242.121,78.121)\Line(237.879,78.121)(242.121,73.879)
\Text(248,86)[]{\normalsize{\Black{$\nu$}}}
\Line(237,80)(243,86)\Line(237,86)(243,80)
\Text(200,66)[]{\normalsize{\Black{$+$}}}
\Line[dash,dashsize=0.5](180,76)(180,86)
\Line
[dash,dashsize=0.5,arrow,arrowpos=1,arrowlength=2.5,arrowwidth=1,arrowinset=0.2](183,86)(183,78)
\Text(160,56)[]{\normalsize{\Black{$\mu$}}}
\Line(190,66)(150,66)
\Line(157,63)(163,69)\Line(157,69)(163,63)
\Arc[dash,dashsize=0.5](180,66)(10,-0,180)
\COval(180,86)(4.243,4.243)(45.0){Black}{White}\Line
(177.879,83.879)(182.121,88.121)\Line(177.879,88.121)(182.121,83.879)
\Text(142,66)[]{\normalsize{\Black{$-$}}}
\Text(188,96)[]{\normalsize{\Black{$\nu$}}}
\Line(177,90)(183,96)\Line(177,96)(183,90)
\end{picture}
\label{b-19}%
\end{equation}
The integrals for the first two diagrams in the above cancel each other after
loop momentum shifting which is allowed under dimensional regularization. The
amplitude for the 3rd diagram also vanishes because the fermion loop that may
produce Levi-Civita tensor terms is essentially embedded in a two point
function that lacks sufficient indices to form a Levi-Civita tensor. The
propagators for the two internal lines that are attached to the fermion line
in the 4th or 5th diagram must both be $D\left(  H,H\right)  $. The amplitude
for this type of diagrams is absent of Levi-Civita tensor term because a
triangular fermion loop with one vector $A$ and two scalar $H$ lines attached
does not have enough indices available to make up a Levi-Civita tensor. By
$\left(  \ref{gbbwti}\right)  $, the 6th diagram in $\left(  \ref{b-19}%
\right)  $ can be decomposed as
\begin{equation}%
\begin{picture}(210,39) (5,-27)
\SetWidth{0.5}
\SetColor{Black}
\Line
[arrow,arrowpos=1,arrowlength=2.5,arrowwidth=1,arrowinset=0.2](100,-8)(94,-8)
\Line
[arrow,arrowpos=1,arrowlength=2.5,arrowwidth=1,arrowinset=0.2](170,-8)(194,-8)
\Line[dash,dashsize=0.5](200,-11)(190,-1)
\Line[dash,dashsize=0.5](190,-1)(180,-1)
\Text(198,2)[]{\normalsize{\Black{$\nu$}}}
\Text(160,-21)[]{\normalsize{\Black{$\mu$}}}
\Line(200,-11)(150,-11)
\Line(187,-4)(193,2)\Line(187,2)(193,-4)
\Line(157,-14)(163,-8)\Line(157,-8)(163,-14)
\Line
[dash,dashsize=2,arrow,arrowpos=0.5,arrowlength=2.5,arrowwidth=1,arrowinset=0.2](180,-1)(170,-11)
\CBox(182.828,-3.828)(177.172,1.828){Black}{Black}
\COval(170,-11)(4.243,4.243)(45.0){Black}{White}\Line
(167.879,-13.121)(172.121,-8.879)\Line(167.879,-8.879)(172.121,-13.121)
\Line[dash,dashsize=0.5](130,-11)(120,-1)
\Line[dash,dashsize=0.5](120,-1)(110,-1)
\Text(140,-11)[]{\normalsize{\Black{$+$}}}
\Text(128,2)[]{\normalsize{\Black{$\nu$}}}
\Text(90,-21)[]{\normalsize{\Black{$\mu$}}}
\Line(130,-11)(80,-11)
\Line(117,-4)(123,2)\Line(117,2)(123,-4)
\Line(87,-14)(93,-8)\Line(87,-8)(93,-14)
\Line
[dash,dashsize=2,arrow,arrowpos=0.5,arrowlength=2.5,arrowwidth=1,arrowinset=0.2](110,-1)(100,-11)
\CBox(112.828,-3.828)(107.172,1.828){Black}{Black}
\COval(100,-11)(4.243,4.243)(45.0){Black}{White}\Line
(97.879,-13.121)(102.121,-8.879)\Line(97.879,-8.879)(102.121,-13.121)
\Line[dash,dashsize=0.5](60,-11)(50,-1)
\Line[dash,dashsize=0.5](50,-1)(40,-1)
\Text(70,-11)[]{\normalsize{\Black{$=$}}}
\Text(58,2)[]{\normalsize{\Black{$\nu$}}}
\Text(20,-21)[]{\normalsize{\Black{$\mu$}}}
\Line(60,-11)(10,-11)
\Line(47,-4)(53,2)\Line(47,2)(53,-4)
\Line(17,-14)(23,-8)\Line(17,-8)(23,-14)
\Line
[dash,dashsize=2,arrow,arrowpos=0.5,arrowlength=2.5,arrowwidth=1,arrowinset=0.2](40,-1)(30,-11)
\CBox(42.828,-3.828)(37.172,1.828){Black}{Black}
\COval(30,-11)(4.243,4.243)(45.0){Black}{White}\Line
(27.879,-13.121)(32.121,-8.879)\Line(27.879,-8.879)(32.121,-13.121)
\end{picture}
\label{b22}%
\end{equation}
The fermion loop for either diagram on the right side of the above identity
has only two vertices effectively and will not have enough indices to give
rise to any Levi-Civita tensor term.

The last diagram in $\left(  \ref{b-19}\right)  $ may be problematic because
its cut point is located next to the composite vertex $%
\begin{picture}(12,12) (10,-8)
\SetWidth{0.5}
\SetColor{Black}
\COval(16,-4)(5.657,5.657)(45.0){Black}{White}\Line
(13.172,-6.828)(18.828,-1.172)\Line(13.172,-1.172)(18.828,-6.828)
\end{picture}%
$ as in $\left(  \ref{wti3}\right)  $ of which the non-vanishing amplitude
invalidates the basic identity $\left(  \ref{gbbwti}\right)  $ to result in
the 1-loop anomaly. Let us recall that a cut point not residing in a divergent
sub-diagram of self-energy insertion or vertex correction is said to be
proper. Since a proper cut point for a generator diagram remains to be a
proper one for any of the component diagrams constructed by attaching $%
\begin{picture}(12,12) (10,-8)
\SetWidth{0.5}
\SetColor{Black}
\COval(16,-4)(5.657,5.657)(45.0){Black}{White}\Line
(13.172,-6.828)(18.828,-1.172)\Line(13.172,-1.172)(18.828,-6.828)
\end{picture}%
$ to the generator, we will choose to cut each generator in $\left(
\ref{a-03-1}\right)  $ at a legitimate and proper point if it is available.
For the seven generators in $\left(  \ref{a-03-1}\right)  $, only the third
diagram does not have such a cut point when the non-fermion internal line
corresponds to a wavy line representing an internal vector meson line. But
this is the situation that we have already encountered in Section \ref{2ltd}
in constructing proper component diagrams from the third generator diagram of
$\left(  \ref{a-02-1a}\right)  $.

To violate a Ward identity in our $\gamma_{5}$ scheme, a cut point must be
positioned next to the $%
\begin{picture}(12,12) (10,-8)
\SetWidth{0.5}
\SetColor{Black}
\COval(16,-4)(5.657,5.657)(45.0){Black}{White}\Line
(13.172,-6.828)(18.828,-1.172)\Line(13.172,-1.172)(18.828,-6.828)
\end{picture}%
$\ vertex, such as the one for the\ last diagram in $\left(  \ref{b-19}%
\right)  $. With $\gamma_{5}$ positioned immediately to the right of the
composite vertex $%
\begin{picture}(12,12) (10,-8)
\SetWidth{0.5}
\SetColor{Black}
\COval(16,-4)(5.657,5.657)(45.0){Black}{White}\Line
(13.172,-6.828)(18.828,-1.172)\Line(13.172,-1.172)(18.828,-6.828)
\end{picture}%
$, the $%
\begin{picture}(12,12) (10,-8)
\SetWidth{0.5}
\SetColor{Black}
\COval(16,-4)(5.657,5.657)(45.0){Black}{White}\Line
(13.172,-6.828)(18.828,-1.172)\Line(13.172,-1.172)(18.828,-6.828)
\end{picture}%
$ vertex together with the double line pointing to the right in the identity
$\left(  \ref{gbbwti}\right)  $ is no longer equal to $igR$\ but should be
interpreted as
\begin{equation}
\left(  igR\left(  \not \ell -m\right)  \right)  |_{rightmost\text{ }%
\gamma_{5}}\frac{1}{\left(  \not \ell -m\right)  }=igR-ig\gamma_{5}%
\not \ell _{\Delta}\frac{1}{\left(  \not \ell -m\right)  } \label{b23}%
\end{equation}
The extra term $-ig\gamma_{5}\not \ell _{\Delta}\frac{1}{\left(
\not \ell -m\right)  }$ may contribute to the violation of the Ward identity
and give rise to an anomaly. If we restrict ourselves to legitimate and proper
cut points, only the following four kinds of diagrams, one of which is the
last diagram in $\left(  \ref{b-19}\right)  $, may be responsible for the
violation of the 2-loop triangular Ward identity.%
\begin{equation}%
\begin{picture}(246,19) (9,-6)
\SetWidth{0.5}
\SetColor{Black}
\CBox(52.828,6.172)(47.172,11.828){Black}{Black}
\Line[dash,dashsize=0.5](50,9)(40,-1)
\Line
[dash,dashsize=2,arrow,arrowpos=0.5,arrowlength=2.5,arrowwidth=1,arrowinset=0.2](50,9)(60,-1)
\Line(60,-1)(10,-1)
\Line(27,-4)(33,2)\Line(27,2)(33,-4)
\Line(17,-4)(23,2)\Line(17,2)(23,-4)
\COval(60,-1)(4.243,4.243)(45.0){Black}{White}\Line
(57.879,-3.121)(62.121,1.121)\Line(57.879,1.121)(62.121,-3.121)
\Line(250,-1)(210,-1)
\COval(250,-1)(4.243,4.243)(45.0){Black}{White}\Line
(247.879,-3.121)(252.121,1.121)\Line(247.879,1.121)(252.121,-3.121)
\Line(232,-4)(238,2)\Line(232,2)(238,-4)
\Line[dash,dashsize=0.5](230,9)(220,-1)
\Line[dash,dashsize=0.5](240,9)(230,9)
\Line
[dash,dashsize=2,arrow,arrowpos=0.5,arrowlength=2.5,arrowwidth=1,arrowinset=0.2](240,9)(250,-1)
\Line(227,6)(233,12)\Line(227,12)(233,6)
\CBox(242.828,6.172)(237.172,11.828){Black}{Black}
\Line(190,-1)(140,-1)
\COval(190,-1)(4.243,4.243)(45.0){Black}{White}\Line
(187.879,-3.121)(192.121,1.121)\Line(187.879,1.121)(192.121,-3.121)
\Line(147,-4)(153,2)\Line(147,2)(153,-4)
\Line[dash,dashsize=0.5](170,9)(160,-1)
\Line[dash,dashsize=0.5](180,9)(170,9)
\Line
[dash,dashsize=2,arrow,arrowpos=0.5,arrowlength=2.5,arrowwidth=1,arrowinset=0.2](180,9)(190,-1)
\Line(167,6)(173,12)\Line(167,12)(173,6)
\CBox(182.828,6.172)(177.172,11.828){Black}{Black}
\CBox(112.828,6.172)(107.172,11.828){Black}{Black}
\Line[dash,dashsize=0.5](110,9)(100,-1)
\Line
[dash,dashsize=2,arrow,arrowpos=0.5,arrowlength=2.5,arrowwidth=1,arrowinset=0.2](110,9)(120,-1)
\Line(120,-1)(80,-1)
\Line(87,-4)(93,2)\Line(87,2)(93,-4)
\Line(107,-4)(113,2)\Line(107,2)(113,-4)
\COval(120,-1)(4.243,4.243)(45.0){Black}{White}\Line
(117.879,-3.121)(122.121,1.121)\Line(117.879,1.121)(122.121,-3.121)
\end{picture}
\label{b24}%
\end{equation}
For the theory of $\left(  \ref{Lanf}\right)  $ in which we have added another
fermion field to cancel the 1-loop anomaly, the $\Lambda$ factor of mass
dimension 1 from the solid black box
\begin{picture}(5,5) (0,-1)
\SetWidth{0.5}
\SetColor{Black}
\CBox(5.83,-0.83)(0.17,4.83){Black}{Black}
\end{picture}
, which represents the vertex factor of $c-\bar{c}-H$,\ reduces the power
counting such that the extra term with the $\not \ell _{\Delta}$ factor cannot
survive the $n\rightarrow4$ limit in any diagram of $\left(  \ref{b24}\right)
$. The theory defined by $\left(  \ref{Lanf}\right)  $ is therefore also free
of 2-loop anomaly.

\end{document}